\documentclass{article}

\usepackage{PRIMEarxiv}

\usepackage[utf8]{inputenc} 
\usepackage[T1]{fontenc}    
\usepackage{hyperref}       
\usepackage{url}            
\usepackage{booktabs}       
\usepackage{amsfonts}       
\usepackage{nicefrac}       
\usepackage{microtype}      
\usepackage{lipsum}
\usepackage{fancyhdr}       
\usepackage{graphicx}       
\graphicspath{{media/}}
\usepackage{array}
\usepackage{listings}
\usepackage{xcolor}
\usepackage{orcidlink}

\lstdefinestyle{json}{
  basicstyle=\ttfamily\footnotesize,
  numbers=left,
  numberstyle=\tiny\color{gray},
  stepnumber=1,
  numbersep=5pt,
  stringstyle=\color{blue},
  showstringspaces=false,
  breaklines=true,
  frame=single,
  backgroundcolor=\color{gray!5},
}

\title{\textnormal{Making the Telemetry Pipeline Measurable: An Edge IoT Testbed and Dataset for End-to-End Latency, Provenance, and Energy-Aware Benchmarking}}

\author{
  Dimitris Gkoulis\,\orcidlink{0000-0002-7428-5753} \\
  Harokopio University of Athens \\
  \texttt{gkoulis@hua.gr} \\
}

\begin{document}
\maketitle

\begin{abstract}
Edge-centric IoT systems are increasingly evaluated through sensor values or platform features, yet the telemetry pipeline itself, the sequence of sensing, serialization, message mediation, ingestion, and persistence, is rarely released as a measurable object. This limits reproducible benchmarking of timeliness, provenance, and operational cost, especially in early-stage IoT design where real deployments and high-quality open data are often unavailable. I present a measurement-first edge IoT testbed and dataset that elevate timeliness, provenance, and energy to first-class telemetry dimensions. The testbed operationalizes a role-based event-driven architecture with explicit Source–Broker–Aggregator responsibilities and a schema-governed TelemetryMessage envelope (Avro) that embeds correlation identifiers and a uniform, cross-role vocabulary of stage-level timestamps. To trace mediation effects, I extend an MQTT broker (Eclipse Mosquitto) with a plugin that annotates messages with high-precision ingress/egress timestamps. I release the dataset in a dual representation: (i) replayable raw events as compressed Avro streams and (ii) analysis-ready long-format Parquet tables partitioned for efficient query and benchmarking, generated by a deterministic raw-to-tabular transformation pipeline. The contribution is an integrated engineering artifact and dataset built from established components. The demonstration describes one deployment through latency summaries, measurement-context distributions, and energy aggregates; it does not establish a new attribution method or a controlled performance benchmark. By co-releasing heterogeneous telemetry with pipeline timing and energy signals in both replayable and queryable forms, the artifact provides a basis for reproducible exploratory analysis and subsequent benchmarking studies.
\end{abstract}


\section{Introduction}
\label{section:introduction}

The Internet of Things (IoT) is an increasingly pervasive paradigm in which intelligent things, humans, and software systems are interconnected and interact through sensing, actuation, and networked communication \cite{atzori2010internet,rose2015internet,7030240,madakam2015internet,Li2015,6164978}. The defining capability of these things is not only that they observe and influence the physical world, but also that they continuously transform physical phenomena into digital information that can inform human decision-making or automated control \cite{7030240,atzori2010internet,Li2015,gkoulis2026selfhealing}. As IoT technologies mature, they underpin solutions with substantial technical, social, and economic impact across domains such as environmental monitoring \cite{POLYMENI2022100612}, smart buildings and urban infrastructure \cite{8125369,dankan2020internet}, transportation \cite{rose2015internet}, healthcare \cite{kelly2020internet}, and precision agriculture \cite{KHANNA2019218}. Beyond domain impacts, IoT is also associated with broader economic and productivity effects \cite{espinoza2020estimating} and sustainability-oriented innovation agendas \cite{8125369}. Yet this growth intensifies long-standing engineering challenges rooted in heterogeneity and scale: devices differ in capabilities and protocols; deployments evolve over time; and system-level properties such as robustness, Quality of Service (QoS), interoperability, and openness become decisive for user-centric IoT applications \cite{6164978,7030240,Li2015,madakam2015internet,Gigli2011-sx}.

A persistent practical difficulty is that IoT systems are fundamentally data-driven \cite{atzori2010internet,Li2015,gkoulis2026selfhealing}, while data is often the very resource that is missing at the earliest stages of design and evaluation \cite{6164978}. When practitioners begin with limited resources, potentially without access to “things,” they still need to prototype quickly, understand the physical environment, and validate end-to-end behavior before committing to costly deployments \cite{rose2015internet,7030240}. In such settings, relying on open real-world data is frequently impractical due to quality issues, availability constraints, licensing restrictions, and incompleteness \cite{madakam2015internet, gkoulis2026selfhealing}. Purely synthetic or simulated data can reduce friction and enable early experimentation, especially for time-series workloads \cite{GKOULIS2025101500}; however, synthetic traces rarely capture the operational realities that govern whether an IoT system is usable in practice: the timing behavior of the data-production pipeline, the effects of event mediation, and the energy footprint of continuously running edge nodes \cite{6164978,7030240,Li2015}. Consequently, many evaluations remain either value-centric (focused on sensor readings) or infrastructure-centric (focused on platforms), while the data pipeline itself is treated as an opaque conduit rather than a measurable object \cite{7030240,Li2015}.

This paper argues that, for edge-centric IoT systems, the telemetry pipeline should be treated as a first-class measurement target \cite{7030240,6164978}. Contemporary IoT deployments are commonly organized around event-driven communication, where edge nodes publish observations and downstream components consume, transform, and persist them for analytics and actuation \cite{7030240,Gigli2011-sx,Li2015}. In such architectures, timeliness is not a secondary attribute but a functional requirement: delays, jitter, and tail latencies can directly affect alerting, automation, and decision quality \cite{Li2015, GKOULIS2026102051, gkoulis2026selfhealing}. At the same time, practical edge operation is constrained by energy \cite{6164978,7030240,8125369}. Even when the sensing problem appears simple, the continuous cost of compute, networking, and serialization on small devices determines feasibility and longevity \cite{6164978,Li2015}. Despite these realities, datasets and testbed artifacts often under-specify the provenance of telemetry events, omit stage-level timing that enables end-to-end auditing, and report energy separately from the telemetry produced during the same observation windows \cite{6164978,7030240}.

I address this gap by presenting an edge-focused IoT testbed and a dataset that jointly capture heterogeneous telemetry and the behavior of the telemetry pipeline that produced it, motivated by telemetry requirements found in environmental, agricultural, and health-oriented IoT deployments \cite{POLYMENI2022100612,KHANNA2019218,kelly2020internet}. The core idea is \emph{measurement discipline at the edge}, expressed through three consistent differentiating axes that structure the paper: (i) \textbf{pipeline observability}, by systematically recording role-spanning, stage-level timeliness signals across Source, Broker, and Aggregator; (ii) \textbf{energy alignment}, by co-measuring plug-level energy in a way that can be associated with telemetry production and cadence; and (iii) \textbf{dual representation}, by releasing the artifact as both replayable raw events (Avro) and analysis-ready tabular views (partitioned Parquet). Timeliness metadata is embedded directly into a schema-governed telemetry envelope, enabling message correlation and inspection of recorded provenance within the telemetry envelope \cite{Gigli2011-sx,Li2015}. To support both systems-level replay and data-centric analytics, the dataset is released in a dual representation generated by a deterministic transformation pipeline. This combination yields an artifact that is simultaneously interpretable as an event stream and structured for exploratory analysis and subsequent reproducible evaluation \cite{madakam2015internet,rose2015internet}.

The contribution is an integrated testbed and dataset, rather than a new messaging protocol, latency attribution method, energy attribution model, or provenance mechanism. Its practical value lies in making the following established instrumentation and publication choices available together in a documented artifact. The present deployment measures indoor environmental conditions; it does not validate an agricultural, healthcare, or industrial application.

The contributions are:

\begin{itemize}
    \item A role-based edge testbed architecture that operationalizes event-driven IoT dataflow through explicit Source, Broker, and Aggregator roles, suitable for heterogeneous edge nodes.
    \item A schema-governed telemetry envelope that embeds correlation identifiers and stage-level timeliness metadata, supporting inter-event latency estimates and message-level provenance inspection across pipeline mediation. 
    \item Integrated energy co-measurement via plug-level metering aligned with telemetry production, supporting time-windowed analyses of data volume, pipeline behavior, and whole-node operational energy.
    \item A reproducible dataset publication workflow that releases both raw Avro events and curated, partitioned Parquet tables through a deterministic, bidirectional raw-to-tabular transformation pipeline.
\end{itemize}

The remainder of this paper is organized as follows. Section~\ref{section:relatedwork} situates this work within prior IoT testbeds and measurement-oriented infrastructures. Section~\ref{section:testbedarchitecture} introduces the abstract model and architecture of the proposed testbed. Section~\ref{testbed:implementation} details the implementation, including hardware, software roles, telemetry schema, and timeliness instrumentation. Section~\ref{section:testbeddemonstration} demonstrates the resulting dataset and illustrates how it supports inspection of environmental signals, context distributions, latency behavior, and energy signatures. Section~\ref{section:discussion} discusses implications, limitations, and directions for extension, and Section~\ref{section:conclusions} concludes.

\section{Related Work}
\label{section:relatedwork}

IoT testbeds have been developed across a wide spectrum of scales, goals, and methodological commitments, ranging from application-driven smart-building platforms to city-scale experimentation facilities, remotely accessible academic infrastructures, protocol- and technology-specific measurement rigs, virtualized environments, and reproducible dataset-generation testbeds. This section situates this work within these strands and motivates the specific gap addressed by the presented instrumentation- and dataset-centric contribution.

\paragraph{Design goals and academic testbed requirements.}
A recurring theme in the literature is that academic IoT testbeds are expected to support heterogeneity (devices, sensors, and protocols) while remaining controlled and repeatable. The challenges involved in building such a platform are discussed in the context of a smart-building testbed for academic use, including flexibility for communication research, application use cases, cognition, and security, as well as demonstration-oriented monitoring and visualization workflows \cite{8969009}. Complementing such perspectives, a systematic review of IoT/WSN testbed facilities synthesizes the broader landscape, emphasizing the diversity of facilities and the continuing importance of reproducible experimentation practices, supportive materials, and metric-driven facility design \cite{jsan12030048}. This work aligns with these motivations and operationalizes them through a concrete, edge-focused architecture and a dataset publication workflow that makes the data-production pipeline observable.

\paragraph{Large-scale and city-scale experimentation facilities.}
Several influential testbeds emphasize scale, openness, and multi-layer experimentation in real-world conditions. SmartSantander provides a city-scale facility and architectural blueprint for large-scale IoT experimentation in urban scenarios \cite{SANCHEZ2014217}. City of Things similarly offers a multi-technology smart-city testbed with an integrated approach spanning network infrastructure, data experimentation, and living-lab validation \cite{7580875}. FIT IoT-LAB offers a large open experimental platform with thousands of nodes and consistent tooling across multiple sites, enabling experimentation from low-level protocols to higher-level services \cite{7389098}. Work on thorough testbed characterization further highlights how open platforms can support repeatable experimentations and how experimentation can be meaningfully combined with simulation for robust evaluation methodologies \cite{PAPADOPOULOS201786}. In contrast to the primary emphasis of these facilities on scale and broad experimentation enablement, this work's contribution is positioned around \emph{measurement discipline at the edge}: I focus on systematically co-capturing heterogeneous telemetry together with stage-level timing and energy signals, and on releasing analysis-ready artifacts that preserve both replayable raw events and queryable curated views.

\paragraph{Low-cost, minimalistic, and remotely accessible testbeds.}
A complementary direction prioritizes accessibility and ease of deployment using off-the-shelf components. OpenTestBed advances a minimalistic, low-cost design based on commodity hardware and lightweight infrastructure (e.g., MQTT over Wi-Fi) to facilitate adoption across institutions \cite{8845269}. Other platforms emphasize remote access and experiment-as-a-service models, enabling users to deploy and run experiments through remote interfaces and managed resources \cite{8620157}. LinkLab extends this notion by integrating remote development capabilities (web-based IDE, remote compilation) to accelerate iterative development and experimentation across heterogeneous devices \cite{9097619}. This work's implementation similarly leverages off-the-shelf edge nodes and MQTT-based communication, but differentiates itself by treating provenance- and timeliness-oriented instrumentation, together with energy metering alignment, as primary dataset features rather than auxiliary operational concerns.

\paragraph{Technology- and protocol-oriented measurement testbeds.}
A substantial body of work designs testbeds to experimentally characterize communication technologies and quantify key performance parameters. A Universal Testbed is proposed to measure latency, error rate, and stability across multiple wireless IoT technologies while abstracting measurement from specific protocol and hardware choices \cite{s22114159}. Likewise, protocol- and deployment-focused testbeds, such as a low-cost LoRaWAN infrastructure, provide extensive measurement studies that characterize coverage and parameter sensitivity in indoor and outdoor environments \cite{8355180}. These efforts motivate careful measurement practice and produce valuable technology-facing insights. This work is complementary in scope: rather than centering a single wireless technology, I instrument the \emph{end-to-end telemetry pipeline} (Source--Broker--Aggregator) and release data structures that make stage-to-stage timing, message correlation, and energy co-measurement available for downstream benchmarking, tracing, and provenance analyses.

\paragraph{Virtualized and emulated IoT environments.}
Beyond physical infrastructures, virtual testbeds provide controlled ways to emulate tiered IoT architectures and resource constraints. IOTier transforms clusters into a virtual testbed that exposes the performance implications of multi-tier IoT environments under programmable constraints (CPU, memory, storage, network) \cite{9499588}. Emulation-oriented approaches are also central in security-focused platforms such as Gotham, which provides a reproducible and extensible testbed for IoT security experiments and dataset generation across multiple protocols and threat actors \cite{10049670}. This work shares with these systems a commitment to reproducibility and dataset utility, while focusing on a different measurement target: fine-grained, role-spanning timeliness metadata and operational energy signals captured alongside heterogeneous edge telemetry, with a dual raw/tabular release that supports both message-level replay and data-centric analytics.

\paragraph{Energy-aware and batteryless IoT testbeds.}
Energy is a foundational concern in IoT evaluation, particularly for constrained and intermittently powered devices. Shepherd introduces a portable testbed for batteryless IoT that records and replays synchronized energy traces at high resolution to study behavior under realistic spatio-temporal energy availability \cite{ShepherdGeissdoerfer}. The presented testbed addresses a complementary energy perspective aligned with practical edge deployments: plug-level metering captures upstream consumption of compute-and-sense nodes (including adapter effects) and is co-published with telemetry and timing metadata, enabling joint analyses that relate data production, timeliness, and operational energy usage.


\paragraph{Synthesis and gap addressed by this paper.}
Across these directions, prior work has collectively established: (i) the importance of flexible and repeatable academic testbeds \cite{8969009,jsan12030048}; (ii) the value of open, large-scale and city-scale facilities for real-world experimentation \cite{SANCHEZ2014217,7580875,7389098,PAPADOPOULOS201786}; (iii) accessible and remotely usable platforms that lower barriers to experimentation and development \cite{8845269,8620157,9097619}; (iv) rigorous measurement infrastructures for wireless performance characterization \cite{s22114159,8355180}; (v) controlled virtual and reproducible emulation environments, including security-oriented dataset generation \cite{9499588,10049670}; and (vi) energy-centric tooling for constrained/batteryless IoT \cite{ShepherdGeissdoerfer}. 

A useful lens for positioning the contribution of this paper is the same three-axis framing used in the Introduction (Section~\ref{section:introduction}): \textbf{pipeline observability}, \textbf{energy alignment}, and \textbf{dual representation}. The space that remains comparatively under-emphasized is the \emph{co-release of datasets that make the telemetry pipeline itself measurable} at fine granularity, while simultaneously aligning operational energy signals with telemetry production and providing both replayable and analytics-friendly representations. The paper targets this gap by (a) defining a role-based edge architecture with explicit stage instrumentation, (b) embedding timeliness metadata directly into a schema-governed telemetry envelope, (c) incorporating broker-side timestamping to support traceability across message mediation, (d) integrating plug-level energy co-measurement aligned with telemetry cadence, and (e) publishing a dual representation (raw Avro and partitioned Parquet) via a deterministic, reproducible transformation pipeline. This positioning frames the contribution as a measurement-first testbed and dataset. The comparison above is qualitative: it identifies differences in emphasis and artifact organization, not measured advantages in timing accuracy, energy attribution, storage efficiency, or scalability. The instrumentation follows established observability practices; no claim is made that existing testbeds or tracing systems cannot provide comparable measurements. Relative to a value-only sensor trace, the additional fields permit message-level inspection of inter-stage timing and time-windowed association with node energy. Their co-release reduces the need to reconstruct these relationships from separate logs, but the present demonstration does not establish superior diagnostic capability.

\section{Testbed Architecture}
\label{section:testbedarchitecture}

\subsection{Abstract Model}
\label{subsection:abstractmodel}

\begin{figure}[htbp]
  \centering
  \includegraphics[width=0.9\textwidth]{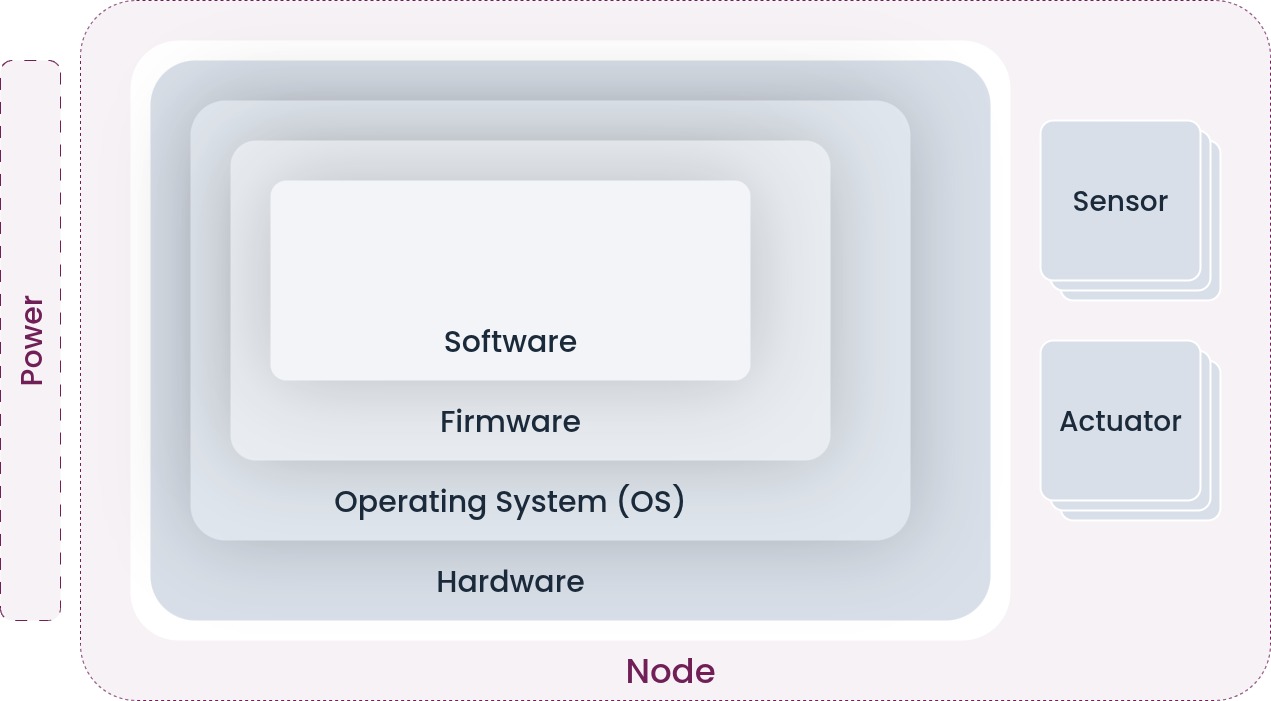}
  \caption{Layered view of an abstract Edge Node, showing hardware, operating system, firmware, software, required power, and optional sensor/actuator extensions.}
  \label{fig:tt-p1}
\end{figure}

\begin{figure}[htbp]
  \centering
  \includegraphics[width=0.9\textwidth]{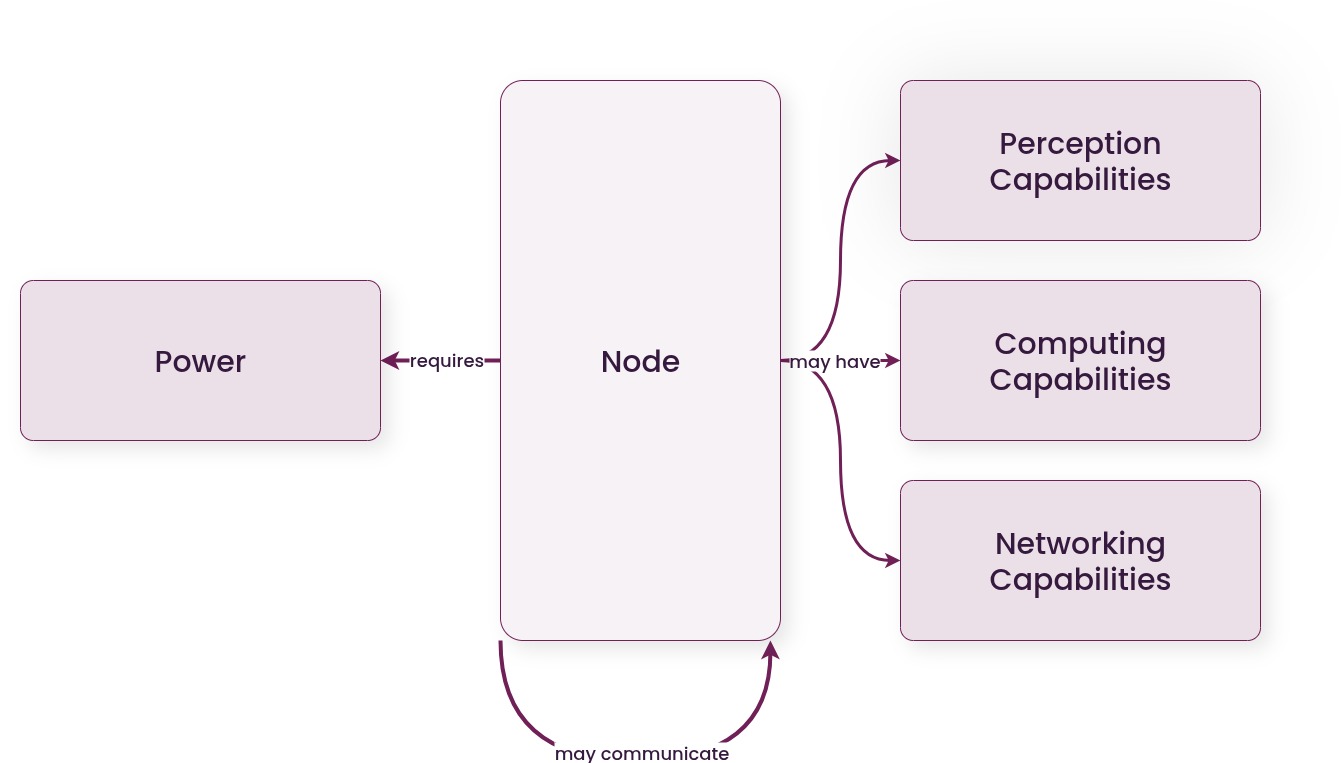}
  \caption{Capability-based abstract model of a Node, showing its power requirement, possible perception, computing, and networking capabilities, and its ability to communicate.}
  \label{fig:tt-p2}
\end{figure}

Based on commonly referenced IoT architectural frameworks in prior studies \cite{8005876, Sobin2020, fi15040135, gkoulis2026selfhealing}, here I provide the perceived abstract model of a Node (Figure~\ref{fig:tt-p1}) as a fundamental component when designing and implementing IoT solutions. The cloud and fog layers are outside the scope of this work. The layer is the Edge. In the Edge there are one or more Nodes.

For a device to qualify as an Edge Node, it must provide at least one of the following (not mutually exclusive) capabilities (Figure~\ref{fig:tt-p2}): 

\begin{itemize}
    \item a unit with \textbf{Perception} (sensing and/or actuating) capabilities—not to be confused with a sensor or actuator, as sensor and actuator are extensions of this kind of unit; imagine an RPI with a sensor attached; 
    \item a unit with \textbf{Computing} capabilities (usually including serialization, deserialization, transformation, and networking; it usually runs an OS and provides good ergonomics so custom software or firmware can be installed), with the understanding that these computation capabilities are always enhanced within the constraints that Edge may involve; 
    \item or a unit with \textbf{Networking} (gateway/router/bridge) capabilities, which is usually a local gateway for, e.g., star topologies, providing access to backhaul networks (Cloud/WAN).
\end{itemize}

In all cases, it must be able to communicate either with other Nodes or with a Node that is a Gateway Unit.

A Node requires power. Power is usually supplied by energy harvesting, battery, or mains.

\subsection{Abstract Testbed Architecture}

\begin{figure}[htbp]
  \centering
  \includegraphics[width=1.0\textwidth]{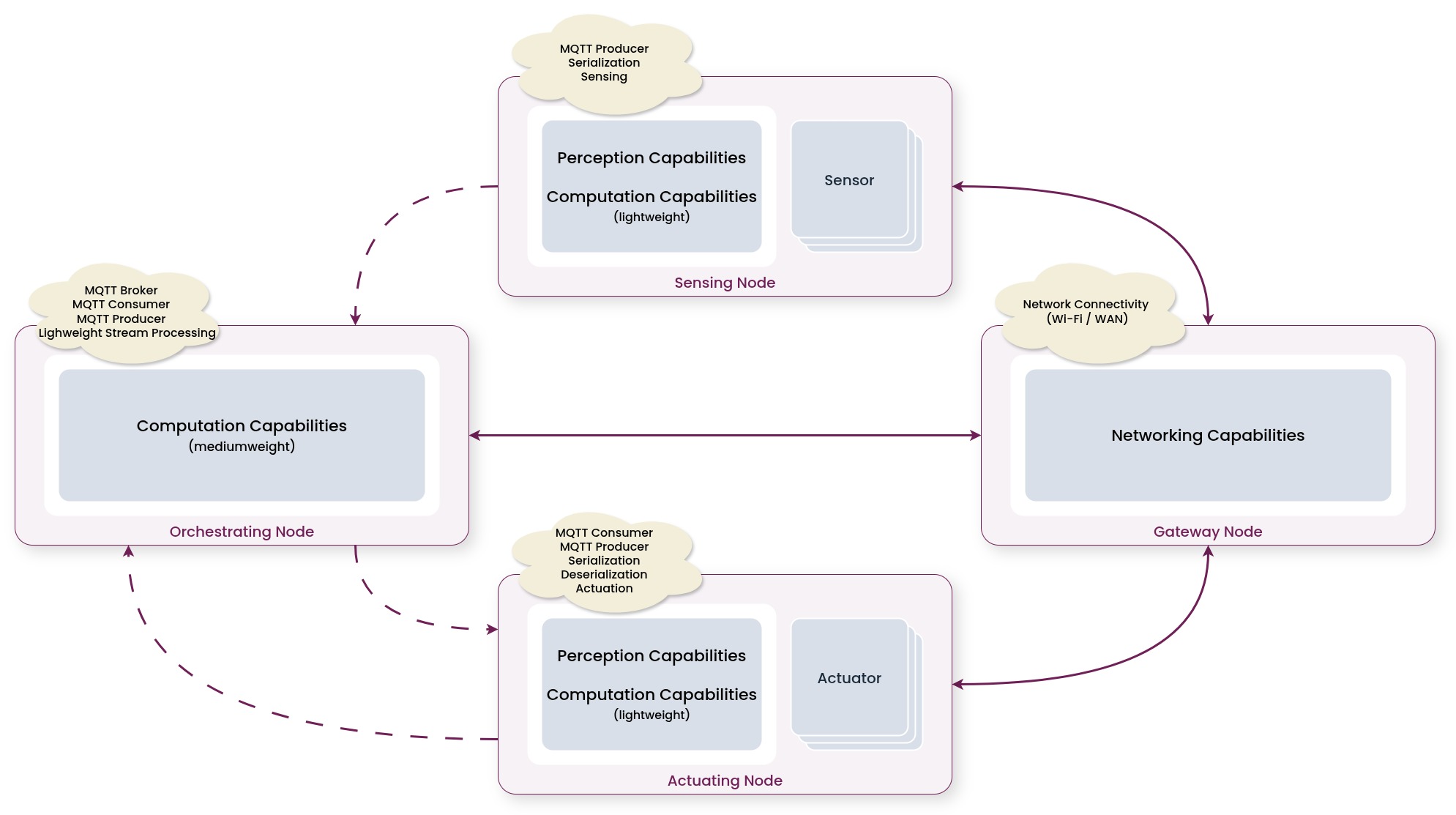}
  \caption{Abstract, technology-agnostic IoT testbed architecture composed of sensing, actuating, orchestrating, and gateway nodes, with MQTT-based communication and networking relationships among them.}
  \label{fig:tt-p3}
\end{figure}

Here, I present an abstract, technology-agnostic testbed architecture. It is derived from the architectures mentioned in Sections~\ref{section:introduction},~\ref{section:relatedwork},~\ref{subsection:abstractmodel}, built using the Node abstract model as the fundamental building block, and adapted to serve common testbed settings with the goal of inspecting data produced from various nodes enriched with energy, timeliness, and provenance data. While a typical IoT solution is more than data (e.g., automation, autonomous, security, and many more), data enriched with energy, timeliness, and provenance information are, in my opinion, the base of a testbed and the purpose it serves because, when structured and documented properly, it can represent the state of a complex, heterogeneous, interconnected system such as IoT in a deterministic, interpretable, replayable manner. Consequently, it help practitioners understand various specifics and nuances at a very early stage of the IoT design and development lifecycle. That said, next I present the architecture components as well as a graphical representation depicted in Figure \ref{fig:tt-p3}. As an example, I include one node by kind to make the example complete.

There is a Node with networking capabilities that acts as the local network’s router and gateway. Specifically, all nodes communicate with each other using this node at a networking level. For example, it can be a Wi-Fi/Ethernet router with some name-resolving capabilities and other common networking functionality.

There is a Node with computational capabilities that acts as a broker of high-level messages in the context of an event-driven approach through appropriate software like MQTT. It provides two communication paradigms, publish–subscribe and request–reply. These two paradigms can serve a plethora of communication scenarios, especially publish–subscribe, which is very common in IoT, even for command/query patterns on top of that.

There is a Node with Perception capabilities, and more specifically sensing capabilities. It senses the environment in various ways based on the sensors attached, either wired or wireless. For instance, it can measure air quality, climate quality, vision, sound, and many more. It usually follows a flow that involves sensing, measuring, transforming, and then sending the data to one or multiple receivers. The most common way is the publish–subscribe paradigm, in which the node publishes the data, appropriately serialized, to a specific topic so one or multiple subscribers can consume the data in near real time.

There is a Node with Perception capabilities, and more specifically actuating capabilities. It is connected, either wired or wireless, with one or more actuators that produce force, torque, or displacement when they receive an appropriate electrical input. That way, automatic control of real things is achieved. Some examples include production line automation, home automation, and more. It usually follows a flow of listening to events/messages and then, when one or more messages arrive, it invokes a sequence of deserializing, transforming, and validating the data and probably initiates an actuation signal. The most common way is the publish–subscribe paradigm, in which the node subscribes to a topic, usually fine-grained and conceptually owned by this node’s identity or role, and then invokes the aforementioned flow to actuate.

There is a Node with Computing capabilities. Specifically, it runs user software on top of a platform that provides orchestration, flow-based operations, event-driven sourcing, and similar approaches. For instance, alerting or automating a vent opening based on temperature and humidity can be translated to a user-defined function that runs in that platform or framework to provide IoT orchestration of nodes with sensing/actuating capabilities. Other common cases involve aggregation, data fusion, lightweight computation, lightweight machine learning, and many more. Again, like the previous, it utilizes the publish–subscribe paradigm and is usually both conceptually and technically a consumer and a producer: a consumer to listen to sensing information and a producer to trigger actions or alerts based on sensing information and a state machine.

\section{Testbed Implementation}
\label{testbed:implementation}

\subsection{Hardware}

The testbed comprises seven edge nodes: three general-purpose computing–and–perception nodes, three perception-focused smart plugs, and one networking gateway.

Three Raspberry Pi 4 single-board computers provide both computation and sensing capabilities. Each unit runs a Linux-based operating system and exposes GPIO and I²C interfaces for peripheral integration. The first Raspberry Pi is interfaced via GPIO with two digital temperature-and-humidity sensors (DHT11 and DHT22). The second hosts an Enviro expansion board that integrates, among other devices, a BME280 sensor (temperature, relative humidity, and barometric pressure) and an LTR-559 sensor (ambient light and proximity). The third is connected via the I²C bus to an MCP9808 high-accuracy digital temperature sensor. These platforms perform local acquisition, lightweight preprocessing, and publication of measurements, and are capable of executing user-defined software for data handling and coordination.

Perception of electrical load and basic actuation are provided by three Shelly Plug S devices. These Wi-Fi smart plugs offer limited on-device computation relative to the Raspberry Pi nodes but supply switching control (on/off) for connected loads and integrated energy metering (power, voltage, and current). They support local automation and scheduling, operate over IEEE 802.11 b/g/n (2.4 GHz), and include protective functions such as over-temperature, overload, and over-current safeguards. The plugs can be addressed locally or via cloud connectivity, depending on deployment preferences.

Network connectivity for the testbed is supplied by a Sercomm H300S residential gateway. This device consolidates broadband access (e.g., VDSL/DSL/fibre, subject to the service provider), Ethernet switching, and dual-band Wi-Fi 5 wireless networking, and additionally offers telephony (VoIP), USB sharing, and local network management. All edge nodes attach to this gateway for intra-testbed communication and backhaul access.

Power distribution follows a controlled topology intended to support energy measurement. Each Raspberry Pi 4 is powered via its AC/DC adapter, which is plugged into a dedicated Shelly Plug S; each Shelly Plug S is then connected to the mains supply. This arrangement enables per-node monitoring of real-world electrical consumption, including the upstream adapter and any conversion losses. The Sercomm H300S gateway is powered from the mains through its vendor-supplied power adapter.

Under this topology, the three Raspberry Pi nodes measure indoor environmental conditions through their attached sensors, execute user-defined software, and communicate with one another via the Sercomm H300S. The three Shelly Plug S devices also connect through the Sercomm H300S and report the power drawn by the corresponding Raspberry Pi systems (inclusive of power-adapter effects). The subsequent Software section details the roles, functions, and architectural elements that orchestrate these hardware components.

\clearpage

\begin{table}[p]
\centering
\vspace*{\fill}
\begin{tabular}{|>{\centering\arraybackslash}m{5cm}|>{\centering\arraybackslash}m{4cm}|}
\hline

\vspace{2pt}\includegraphics[width=3cm]{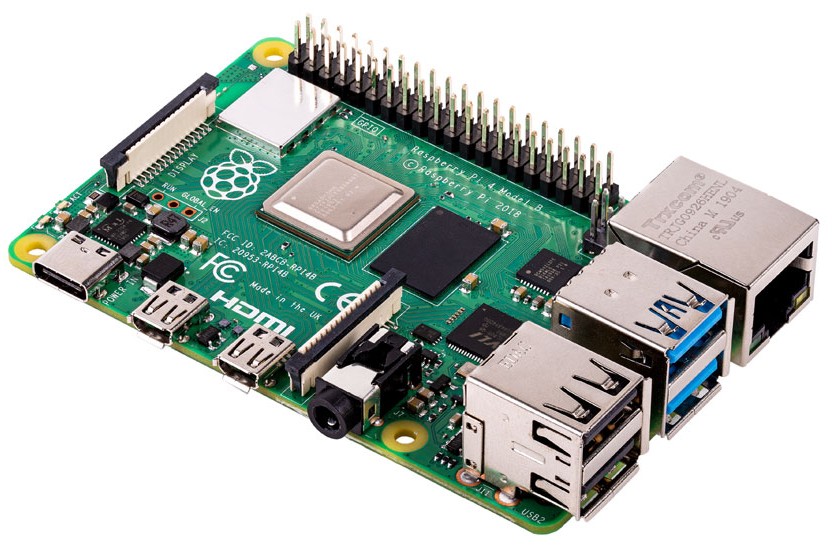} & \textbf{Raspberry Pi 4 single-board computer} \\ \hline

\vspace{2pt}\includegraphics[height=1.4cm]{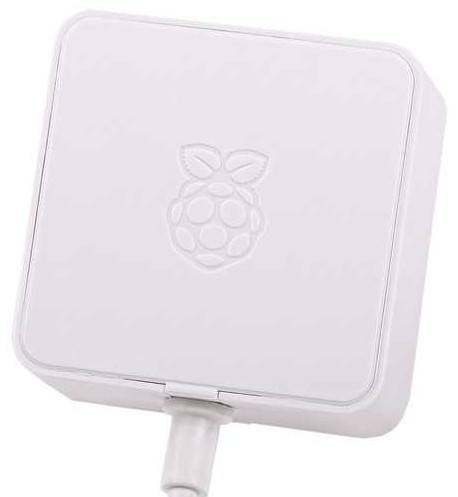} & \textbf{Raspberry Pi 4 Power Supply} \\ \hline

\vspace{2pt}\includegraphics[height=1.6cm]{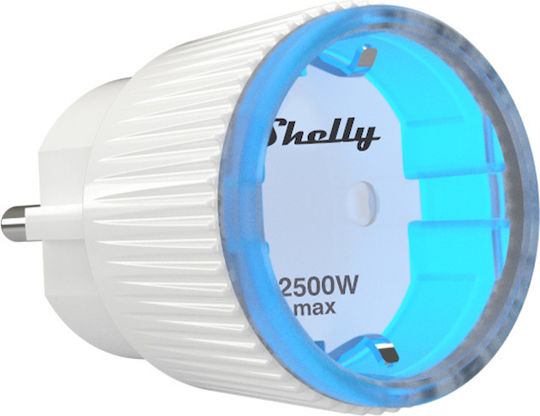} & \textbf{Shelly Plug S device} \\ \hline

\vspace{2pt}\includegraphics[height=1.8cm]{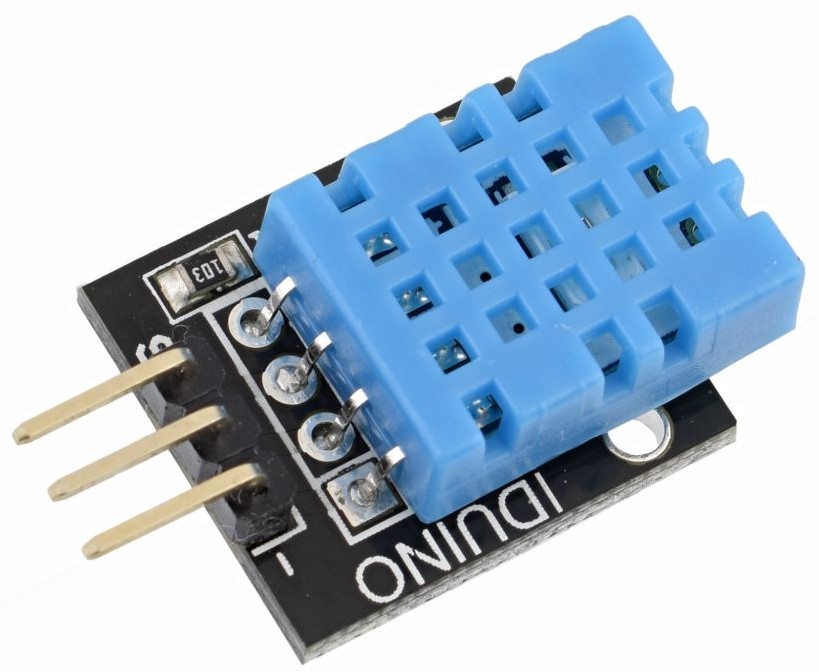} & \textbf{DHT11 Sensor} \\ \hline

\vspace{2pt}\includegraphics[height=1.8cm]{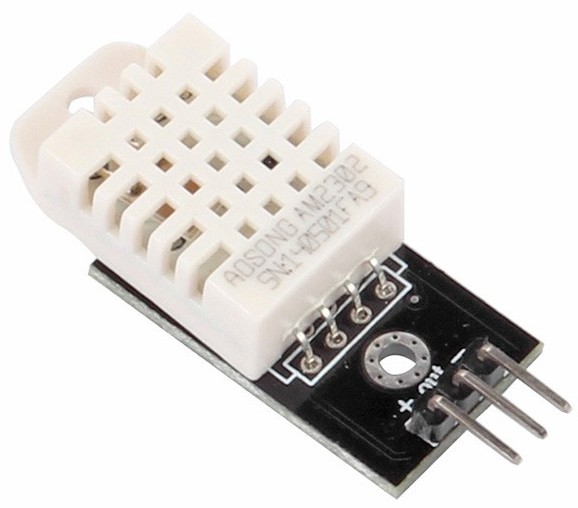} & \textbf{DHT22 Sensor} \\ \hline

\vspace{2pt}\includegraphics[height=1.8cm]{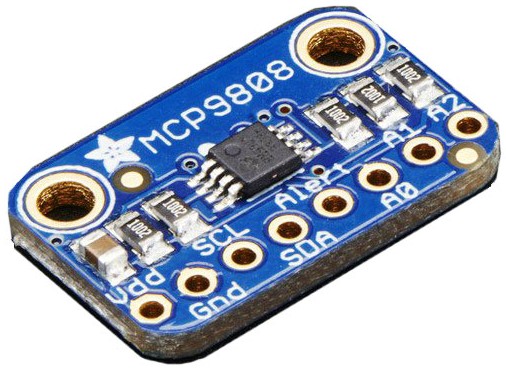} & \textbf{MCP9808 Sensor} \\ \hline

\vspace{2pt}\includegraphics[height=1.8cm]{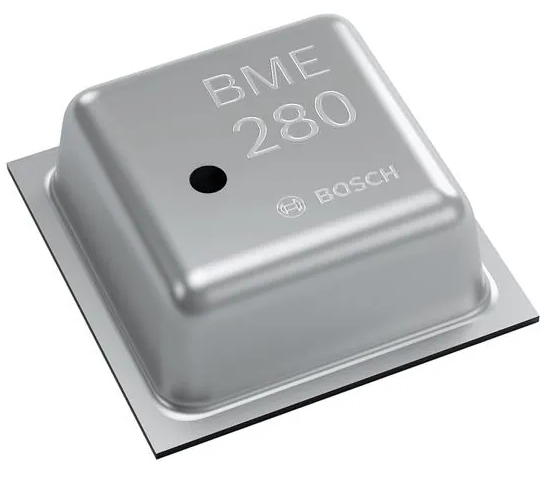} & \textbf{BME280 Sensor} \\ \hline

\vspace{2pt}\includegraphics[height=1.8cm]{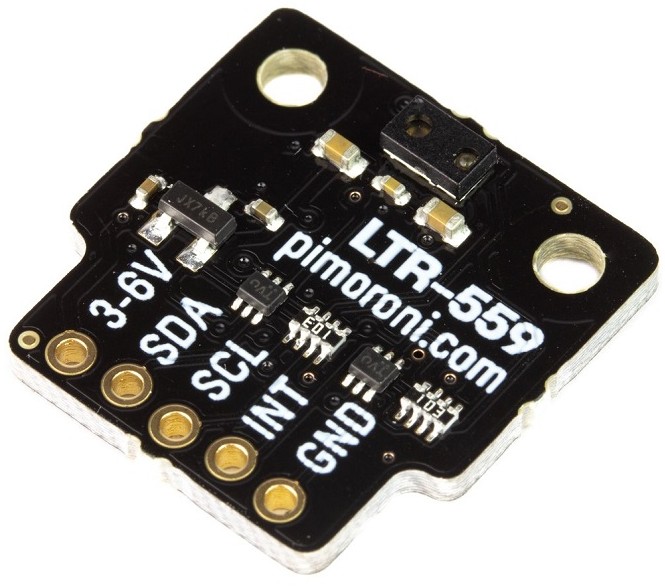} & \textbf{LTR-559 sensor} \\ \hline

\vspace{2pt}\includegraphics[width=2.6cm]{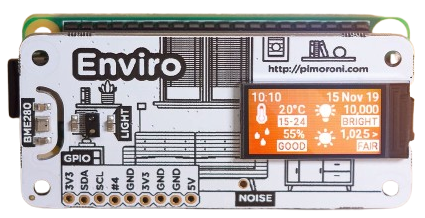} & \textbf{Enviro expansion board} \\ \hline

\vspace{2pt}\includegraphics[height=1.6cm]{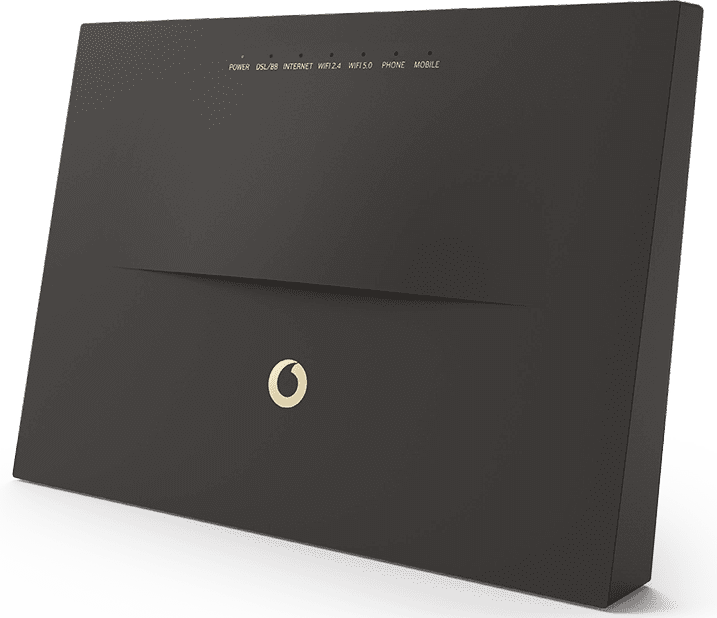} & \textbf{Sercomm H300S residential gateway} \\ \hline

\end{tabular}
\vspace*{\fill}
\caption{Hardware components of the testbed.}
\label{tab:hardware_components}
\end{table}

\clearpage

\subsection{Software}

At a high level, the testbed defines three runtime roles—\textbf{Source}, \textbf{Aggregator}, and \textbf{Broker}—that exchange (produce and consume) \textbf{TelemetryMessage} instances. The Source produces data with explicit timeliness, and provenance semantics from (i) indoor environment sensors, (ii) node/OS condition monitors (e.g., CPU, memory), and (iii) energy meters for computing/perception nodes. The Aggregator consumes all telemetry, persists it locally, and periodically exports curated snapshots to a public fog/cloud repository, thereby materializing the dataset. The Broker mediates event-driven communication.

\subsubsection{Telemetry Message (schema)}

The testbed employs a generic Avro schema for a telemetry envelope used in runtime, in transit, and at rest. It supports one or more readings across one or more contexts per device and embeds detailed timeliness information.

\begin{lstlisting}[style=json]
  {
    "namespace": "testbed0",
    "type": "record",
    "name": "Reading",
    "fields": [
      {
        "name": "value",
        "type": [ "null", "double", "float", "int", "long", "boolean", "string", "bytes" ]
      }
    ]
  },
  {
    "namespace": "testbed0",
    "type": "record",
    "name": "TelemetryMessage",
    "fields": [
      { "name": "correlation_id", "type": "string" },
      { "name": "device_id",      "type": "string" },
      { "name": "timestamp",      "type": "long"   },
      {
        "name": "contexts",
        "type": {
          "type": "map",
          "values": { "type": "map", "values": "testbed0.Reading" }
        }
      },
      {
        "name": "timestamps",
        "type": { "type": "map", "values": "long" }
      }
    ]
  }
\end{lstlisting}

\textbf{Reading} A \textit{Reading} is a polymorphic union holding a single value that may be null, textual (string, bytes), boolean, or numeric (int, long, float, double). It represents a sensed/measured/transformed value.

\textbf{Contexts} A device may expose multiple contexts—physical sensors, virtual sensors, or software subsystems. For example, a Raspberry Pi may expose a DHT11 context with temperature and humidity readings, an OS context with readings such as cpu\_load and mem\_free, and a user-defined virtual context providing a derived indoor-condition risk index from temperature and humidity.

\textbf{Device identity} \textit{device\_id} is a free-form string. In this testbed, Raspberry Pi nodes use their serial number (prefixed to denote type), while Shelly Plug S devices use their MAC address without colons (again with a type prefix). If a device exposes multiple MAC addresses, the active one is used. If primary network interfaces change frequently, the serial number convention is preferred. Failure to retrieve a serial number is treated as an error. Where feasible, identifiers may be anonymized via a hash function; Shelly devices cannot change the internal device ID, but their MQTT topic prefix can be configured (and manually hashed) to enhance privacy.

\textbf{Correlation}. \textit{correlation\_id} is a domain-defined string used to correlate two or more messages (e.g., those belonging to a single sensing cycle). In this implementation it is a UUIDv4.

\textbf{Timeliness}. \textit{timestamp} and the \textit{timestamps} map carry scheduled and stage-level timing metadata (see “Timeliness” below) to support latency measurement, tracing, auditing, and provenance analyses.

\subsubsection{Broker}

The system uses MQTT (OASIS standard) as a lightweight publish/subscribe transport. The implementation is Eclipse Mosquitto, deployed on a Raspberry Pi. Mosquitto is extended with a custom plugin that annotates MQTT messages with high-precision timestamps at ingress and egress, facilitating message tracing, auditing, and provenance. \footnote{Source code: \url{https://github.com/gkoulis/mqtt-message-timestamps}} 

\subsubsection{Source}

The Source role follows a chain-of-responsibility pipeline:

initiate → sense \& measure → transform → serialize → publish

\begin{itemize}
    \item Scheduler (\textbf{initiate}). A blocking scheduler (APScheduler) triggers every 5 seconds, aligned to the Unix epoch. On trigger, it initializes the pipeline’s shared state, records the scheduled timestamp, assigns a UUIDv4 correlation ID, and sets additional provenance metadata.
    
    \item \textbf{Sense \& measure}. Via the Device Manager (below), the Source reads all available and live/healthy sensors—physical and virtual—including libraries that abstract physical sensors (e.g., OS metrics). “Sense” encompasses the physical acquisition, while “measure” denotes conversion by the digital sensor into a discrete value (integer/boolean/decimal).
    
    \item \textbf{Transform}. Values are validated and enriched into structured readings. The handler sets the \textit{correlation\_id}, \textit{device\_id}, initializes and populates timestamps (timeliness), and assembles contexts with the per-context reading maps in the Avro-JSON runtime representation (e.g., a numeric 30 becomes { "temperature": { "value": 30 } } interpreted as 30 °C, with an attached timestamp).

    \item \textbf{Serialize}. The message is serialized using the Avro system from Avro-JSON to Avro binary.

    \item \textbf{Publish}. The payload (Avro binary) is published to MQTT under the topic: \newline\textit{testbed0/\{source\_type\}/\{source\_id\}/telemetry}, where \texttt{source\_type} $\in$ \{RaspberryPi, ShellyPlusPlugSV2\}, and \texttt{source\_id} is the non-prefixed device identifier (Raspberry Pi serial number; Shelly MAC without ``:").
\end{itemize}

\subsubsection{Device Manager}

The Device Manager is a flexible library that auto-registers Raspberry Pi sensors. Given a directory with a defined structure and naming convention, each script implements a sensor contract with methods: \textit{initialize}, \textit{supported}, \textit{read}, and \textit{destroy}. Sensors that initialize without exception, report \textit{supported = true}, and can read are registered. The manager offers ergonomics such as a method returning all current sensings/measurements. Some sensors impose minimum read intervals and may return repeated values if polled too frequently; the 5 s interval used here is conservative for this testbed’s sensors.

\subsubsection{Aggregator}

The Aggregator role also follows a chain-of-responsibility pipeline:

initiate → deserialize → transform → serialize → persist

\begin{itemize}
    \item \textbf{Initiate}. An MQTT consumer listens to all telemetry topics in real time. For each message (in arrival order), the pipeline’s shared state is initialized with the raw byte payload.

    \item \textbf{Deserialize}. Using Avro deserialization, the payload is parsed into the runtime TelemetryMessage representation and validated against the schema. Both Avro-JSON and Avro binary encodings are supported.

    \item \textbf{Transform}. The telemetry object is enriched with additional timeliness timestamps and related metadata.

    \item \textbf{Serialize}. The message is serialized again via Avro. This step both refreshes the enrichment and emulates realistic environments where (de)serialization occurs in-line.

    \item \textbf{Persist}. The serialized Avro binary is stored in an embedded database used as an object/time-series store. Concretely, SQLite is employed with an appropriate schema and semantics.
\end{itemize}

In parallel, the Aggregator performs periodic snapshotting. At a user-configurable interval (default hourly), it creates a database snapshot, records the first and last unprocessed telemetry events, and transforms the selected events into a flattened tabular time-series. In memory these are pandas DataFrames; on disk they are stored as Parquet files. The transformation is bidirectional: tabular rows can be reconstituted into TelemetryMessage objects, supporting practitioners who may wish to operate on raw JSON payloads/MQTT messages or on tabular/time-series formats interchangeably.

A dedicated SQLite catalog also records all collection epochs (ingestion windows), enabling efficient indexing of Parquet files that hold the raw data. After export, the ephemeral snapshot is destroyed.

The export workflow performs a zero-copy transfer of the Parquet artifact(s) into a local Git repository, followed by compression and any additional tasks. It then executes common Git commands such as \textit{add}, \textit{commit}, and \textit{push} to store the dataset in the repository, ultimately yielding a public, transparent, reproducible, and provenance-aware publication of the dataset to the fog/cloud repository.

\subsubsection{Additional Details}

On Raspberry Pi nodes, Source and (optionally) Aggregator are provided by a Python library and run as Linux services. All Raspberry Pi nodes may act as Sources; typically, only one node is designated as the Aggregator.

On Shelly Plug S devices, computation is constrained; thus, each Shelly implements only the Source role. The implementation mirrors the Python logic as closely as possible, using JavaScript and Shelly’s device APIs. Scheduling uses JavaScript timers and is best-effort with alignment to epoch boundaries.

\subsubsection{Timeliness}

Timeliness is a first-class concern. The released dataset records named stage events across Source, Broker, and Aggregator roles. As detailed in Section~\ref{subsec:dataset}, these are event timestamps rather than explicit start/end pairs for every stage. The scheduled timestamp represents the intended sensing-cycle time, whereas stage timestamps mark recorded events in the processing path. The interval from the scheduled timestamp to Source initiation therefore includes scheduling delay and should not be interpreted as sensor acquisition time.

Stage-to-stage delays are obtained by subtracting selected event timestamps for the same message along the pipeline order. An interval labeled by its destination event describes the elapsed time since the preceding selected event; it is not, by itself, the execution time of the destination handler. In particular, the interval ending at broker ingress can include upstream publication, transport, and scheduling effects, while the ingress-to-egress interval describes the instrumented broker segment. Queueing and execution cannot generally be separated without additional boundary events. Likewise, a timestamp preceding persistence does not establish durable database-commit completion.

All nodes run an NTP client, with a Raspberry Pi serving as the local NTP reference. This configuration provides a common time base, but independent measurements of residual clock offsets, drift, and effective timestamp resolution are not reported for this capture. Nanosecond storage units and high-resolution timestamp APIs do not establish nanosecond measurement accuracy. Cross-node differences include relative clock error as well as elapsed time, whereas same-clock intervals avoid a cross-device offset but remain subject to clock adjustments and instrumentation effects. The reported latencies should therefore be read as estimates for the recorded event boundaries, without a validated error bound.

\section{Testbed Demonstration}
\label{section:testbeddemonstration}

This section demonstrates how the proposed testbed materializes a dataset that is simultaneously (i) multi-context (environmental sensing, OS telemetry, and energy metering), (ii) time-aligned via explicit stage-level timestamps, and (iii) amenable to inspection in both raw message form and curated tabular form. The goal is to document what the dataset contains and provide a reproducible first-pass exploration of its structure and recorded measurement characteristics. This is a descriptive artifact demonstration, not a controlled characterization of the telemetry pipeline: message rate, payload size, MQTT QoS, subscriber count, broker/CPU load, and network conditions are not systematically varied. The summaries below describe the captured configuration and do not establish causal effects or performance bounds.

\subsection{Dataset}
\label{subsec:dataset}

The testbed produces a multi-modal telemetry dataset that combines (i) environmental measurements, (ii) node/OS monitoring, (iii) electrical power and cumulative energy metering, and (iv) pipeline-stage timestamps for timeliness and provenance. The dataset is published in two complementary representations: a raw event stream (schema-governed) and a processed, analysis-oriented tabular form. Both are derived from the same \textit{TelemetryMessage} envelope described in Section~5.2.1.

\paragraph{Dataset organization and storage formats.}
The released artifact follows a simple layout with a \texttt{raw/} directory containing compressed Avro files and a \texttt{processed/} directory containing partitioned Parquet:
\begin{itemize}
    \item \textbf{Raw (event-level):} \texttt{dataset/raw/TelemetryMessage-Raw-\{t\_start\}-\{t\_end\}.avro.gz}. Each file stores a sequence of Avro binary records compressed with gzip and chunked by fixed collection windows (hourly in the current release). Records are length-prefixed, enabling streaming reads without full decompression to disk.
    \item \textbf{Processed (tabular):} \texttt{dataset/processed/readings\_long/} and \texttt{dataset/processed/stages\_long/}. Both are written in Parquet and partitioned using Hive-style keys \texttt{(date, hour, device\_type)} to support predicate pushdown and efficient partial scans (e.g., by day, hour, or device class).
\end{itemize}

\paragraph{Reproducible raw-to-tabular transformation.}
The transformation from the raw Avro stream to tabular Parquet is deterministic and implemented as a streaming pipeline. Avro records are decoded into Python dictionaries using the fixed Avro schema (\texttt{telemetry-message.avsc}). Each record is then normalized into two \emph{long-format} tables:

\begin{itemize}
    \item \textbf{\texttt{readings\_long}:} one row per \{message, context, metric\}. The \texttt{contexts} map in each \textit{TelemetryMessage} is exploded such that each metric becomes a separate row. The message timestamp is preserved in nanoseconds (\texttt{ts\_ns}) and as a timezone-aware timestamp (\texttt{ts}); additional convenience fields are materialized (\texttt{sec\_bucket}, \texttt{ts\_sec}, \texttt{date}, \texttt{hour}, \texttt{minute}, \texttt{second}) to simplify windowing and joins.
    \item \textbf{\texttt{stages\_long}:} one row per \{message, stage\}. The \texttt{timestamps} map is exploded so that each recorded pipeline stage contributes a single high-resolution timestamp (\texttt{stage\_ts\_ns}, \texttt{stage\_ts}). In the current release, stage keys are \emph{event timestamps} (not explicit start/end pairs); stage-to-stage delays are computed by subtraction between selected stage events.
\end{itemize}

For scalable export, the normalized tables are registered with DuckDB and written via \texttt{COPY ... (FORMAT PARQUET, PARTITION\_BY (date, hour, device\_type))}. This approach yields compact columnar files and avoids materializing intermediate CSVs or monolithic Parquet outputs.

\paragraph{Tabular schema and semantics.}
Both processed tables are intentionally denormalized to minimize friction for downstream analysis.

\begin{itemize}
    \item \textbf{Identity.} \texttt{device\_id} encodes the device class as a prefix (e.g., \texttt{RaspberryPi-...}, \texttt{ShellyPlusPlugSV2-...}); \texttt{device\_type} is derived from this prefix. The dataset contains three devices of each type (six total).
    \item \textbf{Time.} Timestamps are stored as \texttt{TIMESTAMP WITH TIME ZONE}. The dataset report renders these timestamps with a +03:00 offset, while the underlying values represent absolute epoch-based instants suitable for cross-device alignment. Both \texttt{readings\_long} and \texttt{stages\_long} span the same interval (min: 2025-04-24 21:27:45.022+03:00; max: 2025-04-28 01:09:45.023+03:00), covering 4 distinct dates and 71 distinct date-hour partitions.
    \item \textbf{Values.} Readings are represented using a type-aware layout (\texttt{value\_num}, \texttt{value\_text}, \texttt{value\_bool}) with an explicit \texttt{value\_type}. In this capture, all exported readings are numeric (\texttt{value\_type = number}) and no rows have missing values in the active value column.
    \item \textbf{Contexts.} The \texttt{context} field reflects the provenance of measurements (physical sensors, virtual sensors, platform subsystems). Importantly, the apparent case distinction between \texttt{OS} and \texttt{os} is \emph{intentional}: \texttt{OS} originates from Raspberry Pi OS telemetry, while \texttt{os} originates from Shelly device telemetry. The case therefore encodes a platform boundary rather than a labeling inconsistency.
\end{itemize}

\paragraph{Dataset scale (current release).}
The processed dataset comprises 7{,}721{,}273 rows in \texttt{readings\_long} and 2{,}669{,}334 rows in \texttt{stages\_long}. The dataset is dominated by OS- and energy-related contexts (e.g., \texttt{OS}, \texttt{os}, \texttt{energy}, \texttt{power}), while environmental sensing contexts (e.g., \texttt{BME280}, \texttt{LTR559}, \texttt{MCP9808}, \texttt{DHT11/DHT22}) appear at lower frequencies, reflecting both the higher metric cardinality of system telemetry and the heterogeneity of sensing capabilities across nodes.

\vspace{0.25em}
Overall, this organization preserves the raw event semantics needed for provenance and timeliness analyses, while also providing an efficient, query-friendly representation for exploratory analysis and reproducible benchmarking.

\subsection{Dataset exploration}
\label{subsec:dataset-exploration}

\subsubsection{Windowed environmental signals (10-minute aggregates)}
\label{subsubsec:windowed-environment}

To make environmental trajectories visually interpretable at multi-day scale, sensor readings are aggregated into fixed, non-overlapping 10-minute windows. Each row of the resulting table represents a \{device, context, metric, window\} tuple and includes the mean value, within-window standard deviation, and the number of samples contributing to the estimate. A short excerpt of this representation is shown below.

\begin{table}[htbp]
\centering
\footnotesize
\setlength{\tabcolsep}{3pt}
\begin{tabular}{@{}lllllrrr@{}}
\toprule
Device & Context & Metric & Date & Time & Mean & Std & $n$ \\
\midrule
A & BME280 & humidity    & 2025-04-24 & 21:20:00+03:00 & 50.184458 & 0.114849 & 25  \\
A & BME280 & humidity    & 2025-04-24 & 21:30:00+03:00 & 50.073743 & 0.078786 & 120 \\
A & BME280 & humidity    & 2025-04-24 & 21:40:00+03:00 & 50.026067 & 0.072984 & 120 \\
A & BME280 & humidity    & 2025-04-24 & 21:50:00+03:00 & 50.185135 & 0.076907 & 120 \\
A & BME280 & humidity    & 2025-04-24 & 22:00:00+03:00 & 50.161987 & 0.046389 & 120 \\
\midrule
\multicolumn{8}{c}{\ldots} \\
\midrule
B & MCP9808 & temperature & 2025-04-28 & 00:20:00+03:00 & 22.194271 & 0.024030 & 120 \\
B & MCP9808 & temperature & 2025-04-28 & 00:30:00+03:00 & 22.177083 & 0.027277 & 120 \\
B & MCP9808 & temperature & 2025-04-28 & 00:40:00+03:00 & 22.205208 & 0.041460 & 120 \\
B & MCP9808 & temperature & 2025-04-28 & 00:50:00+03:00 & 22.234896 & 0.032406 & 120 \\
B & MCP9808 & temperature & 2025-04-28 & 01:00:00+03:00 & 22.255826 & 0.025824 & 118 \\
\bottomrule
\end{tabular}
\caption{Excerpt of 10-minute windowed environmental measurements. Each row reports the mean, standard deviation, and sample count $n$ for one device, context, metric, and window. Date and time denote the window start. Both devices are Raspberry Pi nodes: A = \texttt{RaspberryPi-100000003c13fab4}; B = \texttt{RaspberryPi-100000003cab5c68}.}
\label{tab:windowed-environment-excerpt}
\end{table}

\begin{figure}[htbp]
  \centering
  \includegraphics[width=0.9\textwidth]{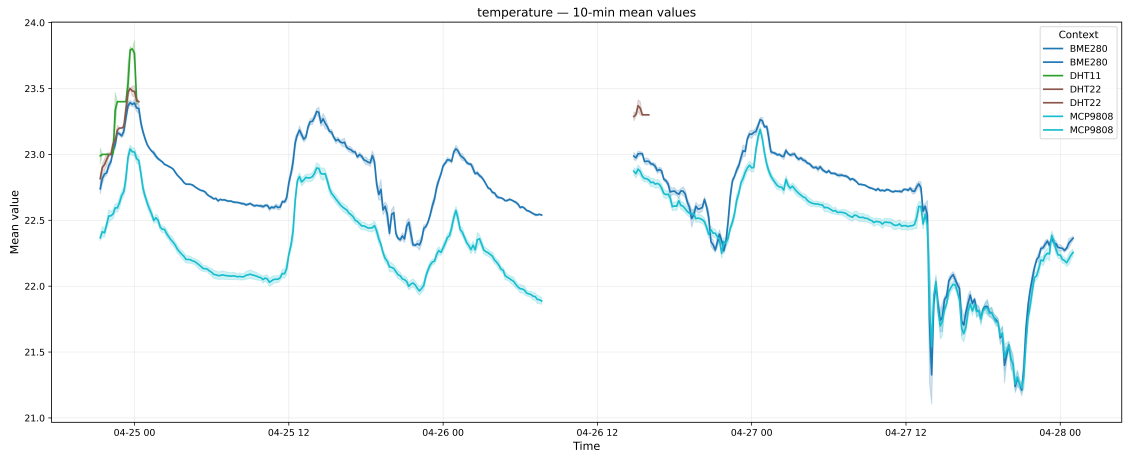}
  \caption{Temperature (10-minute windowed means) across sensing contexts. Segmented traces prevent visual interpolation across missing windows; the windowed representation reduces high-frequency jitter while preserving multi-day trends and enabling cross-context comparison.}
  \label{fig:temp-10min}
\end{figure}

\begin{figure}[htbp]
  \centering
  \includegraphics[width=0.9\textwidth]{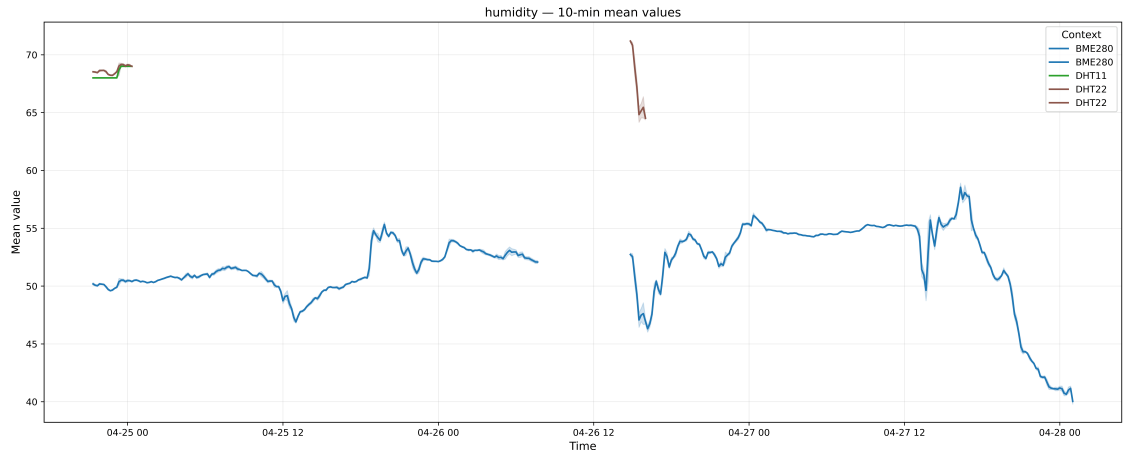}
  \caption{Relative humidity (10-minute windowed means) across sensing contexts. The aggregation exposes slower variations and highlights periods where certain contexts are absent or intermittently reported.}
  \label{fig:humidity-10min}
\end{figure}

\begin{figure}[htbp]
  \centering
  \includegraphics[width=0.9\textwidth]{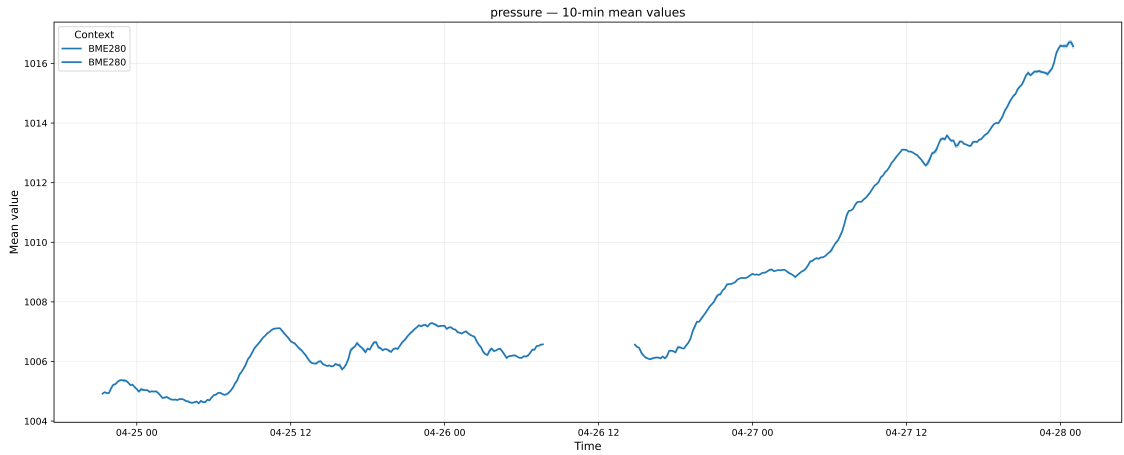}
  \caption{Barometric pressure (10-minute windowed means) for BME280 contexts. Pressure is shown in a windowed form to support multi-day inspection and to avoid over-emphasizing short-term sensor noise.}
  \label{fig:pressure-10min}
\end{figure}

\begin{figure}[htbp]
  \centering
  \includegraphics[width=0.9\textwidth]{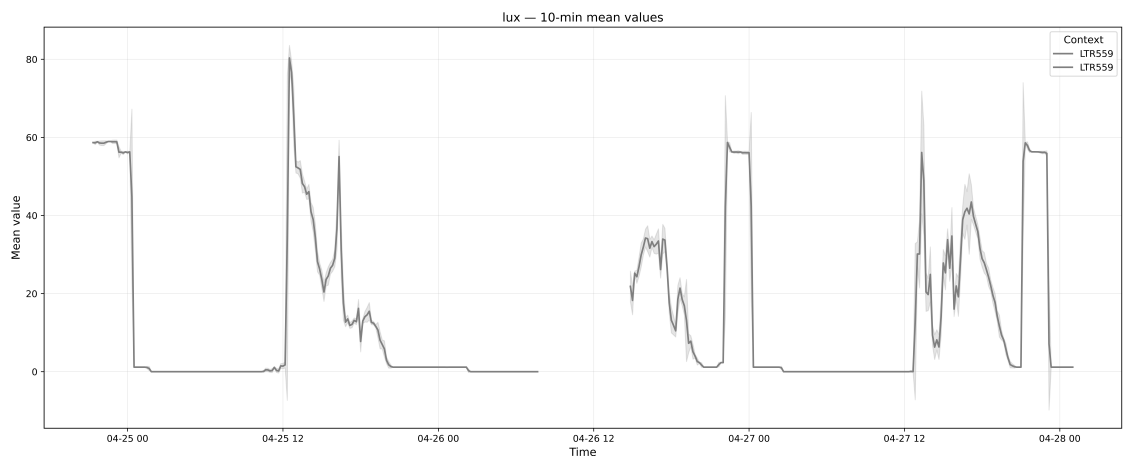}
  \caption{Illuminance (lux; 10-minute windowed means) for LTR559 contexts. The windowed view preserves step-like changes and transient spikes while remaining readable over multi-day spans.}
  \label{fig:lux-10min}
\end{figure}

Figures~\ref{fig:temp-10min}--\ref{fig:lux-10min} visualize these windowed trajectories for temperature, relative humidity, barometric pressure, and illuminance. For each metric, traces are grouped by \textit{context} (e.g., BME280, DHT11/DHT22, MCP9808, LTR559). To avoid implying continuity across missing values or large time discontinuities, traces are \textit{segmented}: gaps in the window sequence are not bridged by line interpolation. Where shown, uncertainty bands are derived from within-window dispersion and the corresponding sample counts, making windows with fewer contributing samples visually distinguishable from fully populated windows.

\textbf{Observed characteristics.} In the demonstrated collection interval, the windowed table contains 2{,}157 aggregated rows spanning approximately 2025-04-24 21:20 through 2025-04-28 01:00 (local time). Sample counts frequently reach $n=120$ per window, consistent with full 10-minute coverage at a nominal 5-second cadence, while boundary windows (e.g., at collection start/end) exhibit smaller $n$. For MCP9808 temperature near the end of the interval, window means cluster tightly (approximately 22.18--22.26$^\circ$C) with low within-window dispersion (std on the order of 0.02--0.04), illustrating stable aggregation under high sample coverage.

\subsubsection{Distribution of measurement contexts}
\label{subsubsec:context-distribution}

\begin{figure}[htbp]
  \centering
  \includegraphics[width=0.9\textwidth]{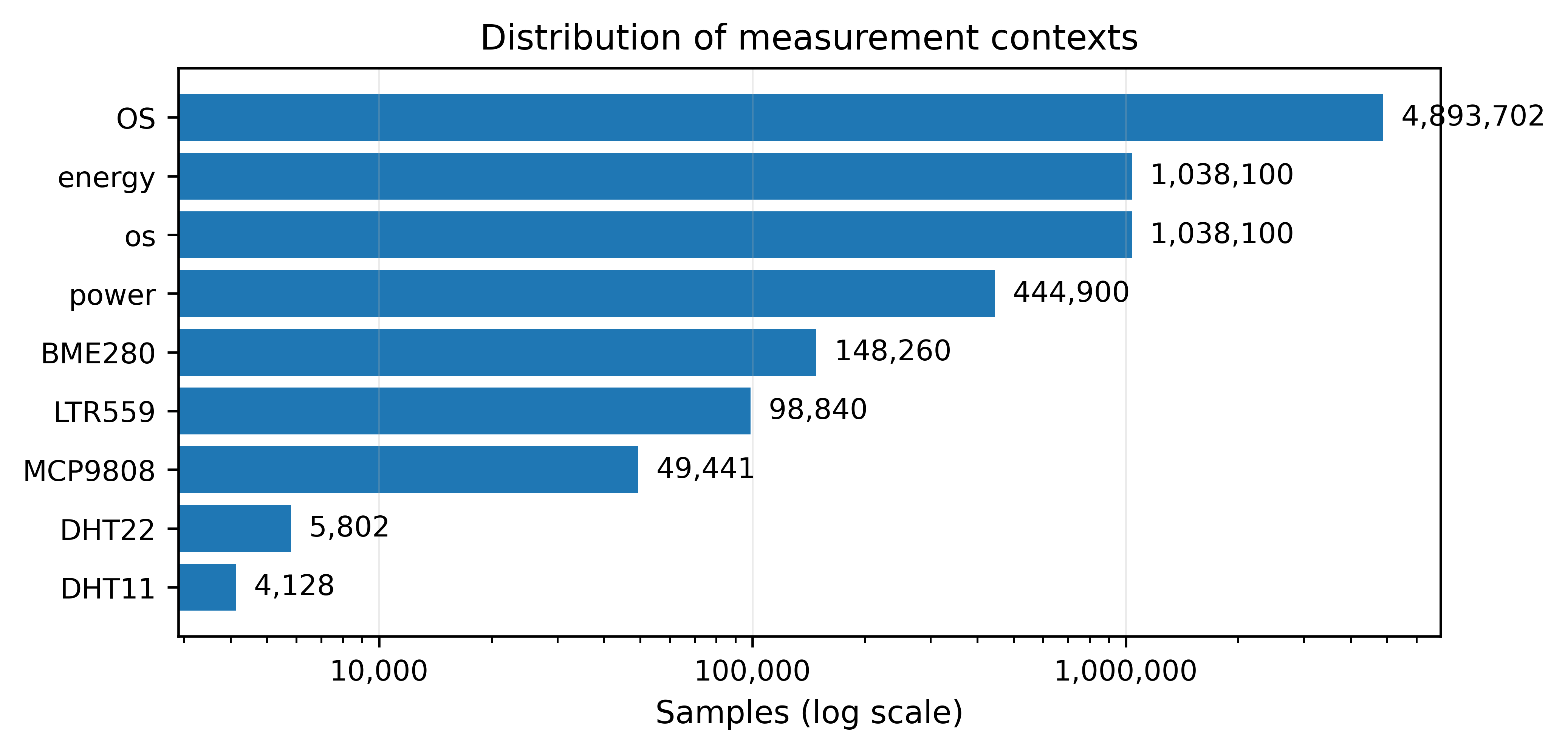}
  \caption{Distribution of measurement contexts (log scale). The dataset is strongly right-skewed: a small number of contexts dominates the total sample volume, while sensor-specific contexts form a visible long tail.}
  \label{fig:context-counts}
\end{figure}

A first compositional check is to quantify how frequently each measurement context appears in the flattened tabular view. Figure~\ref{fig:context-counts} reports the number of recorded samples per context on a logarithmic scale to preserve visibility of low-frequency sensor contexts alongside high-frequency software/energy contexts.

\begin{table}[htbp]
\centering
\setlength{\tabcolsep}{10pt}
\begin{tabular}{lr}
\toprule
Context & Samples \\
\midrule
OS      & 4{,}893{,}702 \\
energy  & 1{,}038{,}100 \\
os      & 1{,}038{,}100 \\
power   &   444{,}900 \\
BME280  &   148{,}260 \\
LTR559  &    98{,}840 \\
MCP9808 &    49{,}441 \\
DHT22   &     5{,}802 \\
DHT11   &     4{,}128 \\
\bottomrule
\end{tabular}
\caption{Sample counts per context in the flattened tabular view (log-scale visualization shown in Figure~\ref{fig:context-counts}).}
\label{tab:context-counts}
\end{table}

\textbf{Observed characteristics.} Across the contexts shown in Table~\ref{tab:context-counts}, the total sample volume is 7{,}721{,}273. The largest category is \texttt{OS} (4{,}893{,}702; 63.4\%), followed by \texttt{energy} (1{,}038{,}100; 13.4\%), \texttt{os} (1{,}038{,}100; 13.4\%), and \texttt{power} (444{,}900; 5.8\%). Environmental sensor contexts appear at lower counts (e.g., BME280 at 148{,}260; LTR559 at 98{,}840). Notably, the distinction between \texttt{OS} and \texttt{os} is intentional: \texttt{OS} originates from Raspberry Pi OS telemetry, whereas \texttt{os} originates from Shelly device telemetry. The case therefore encodes provenance across platforms and should be preserved during grouping and aggregation.

\subsubsection{End-to-end latency by device type}
\label{subsubsec:e2e-latency}

\begin{figure}[htbp]
  \centering
  \includegraphics[width=0.9\textwidth]{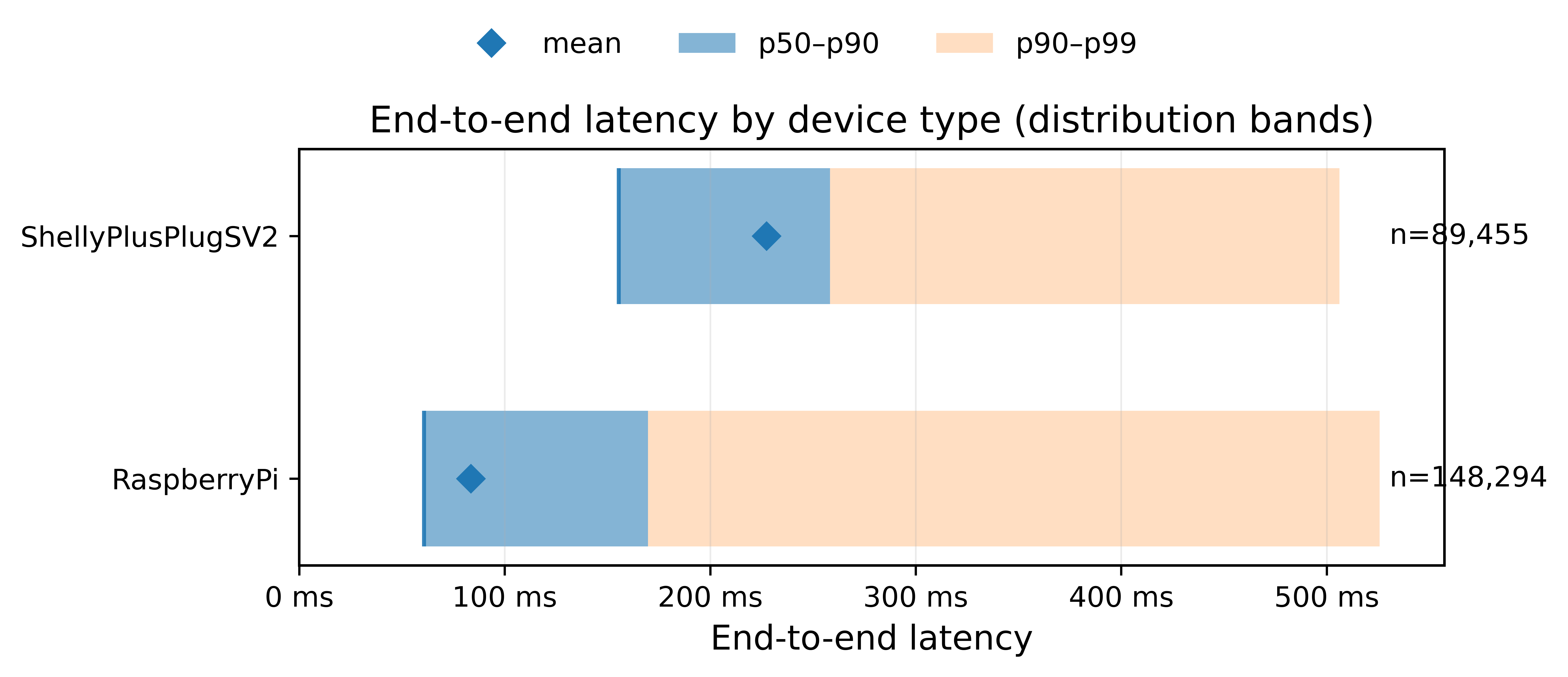}
  \caption{End-to-end latency by device type using percentile bands (p50--p90 and p90--p99) with a mean marker. This representation exposes both typical latency and heavy-tail behavior without collapsing the distribution to a single statistic.}
  \label{fig:e2e-latency}
\end{figure}

Timeliness metadata across Sources, Broker, and Aggregator supports message-level estimates of elapsed time over the recorded pipeline path. Here, end-to-end refers to the instrumented telemetry path, not to a verified physical-sensing-to-durable-storage or sensing-to-actuation delay. Interpretation depends on the selected timestamp endpoints and the clock limitations described above. Figure~\ref{fig:e2e-latency} summarizes these latencies by device type using percentile bands: the inter-percentile span p50--p90 represents the central mass of the distribution, while p90--p99 visualizes tail behavior. A mean marker is overlaid to provide a familiar central-tendency reference.

\begin{table}[htbp]
\centering
\setlength{\tabcolsep}{8pt}
\begin{tabular}{lrr}
\toprule
Metric & RaspberryPi & ShellyPlusPlugSV2 \\
\midrule
avg (ms)  &  83.5068 & 227.3750 \\
p50 (ms)  &  60.6630 & 155.4680 \\
p90 (ms)  & 170.1380 & 258.5950 \\
p95 (ms)  & 237.5740 & 307.0890 \\
p99 (ms)  & 526.2300 & 532.3230 \\
messages  & 148{,}294 &  89{,}455 \\
\bottomrule
\end{tabular}
\caption{End-to-end latency summary by device type (distribution visualization in Figure~\ref{fig:e2e-latency}).}
\label{tab:e2e-latency}
\end{table}

\textbf{Observed characteristics.} The estimated mean end-to-end latency is 227.38~ms for \texttt{ShellyPlusPlugSV2} and 83.51~ms for \texttt{RaspberryPi}. Medians (p50) are 155.47~ms and 60.66~ms, respectively, while p99 exceeds 500~ms for both device types (Table~\ref{tab:e2e-latency}).

\subsubsection{Stage-to-stage latency breakdown (RaspberryPi)}
\label{subsubsec:stage-breakdown}

\begin{figure}[ht]
  \centering
  \includegraphics[width=0.99\textwidth]{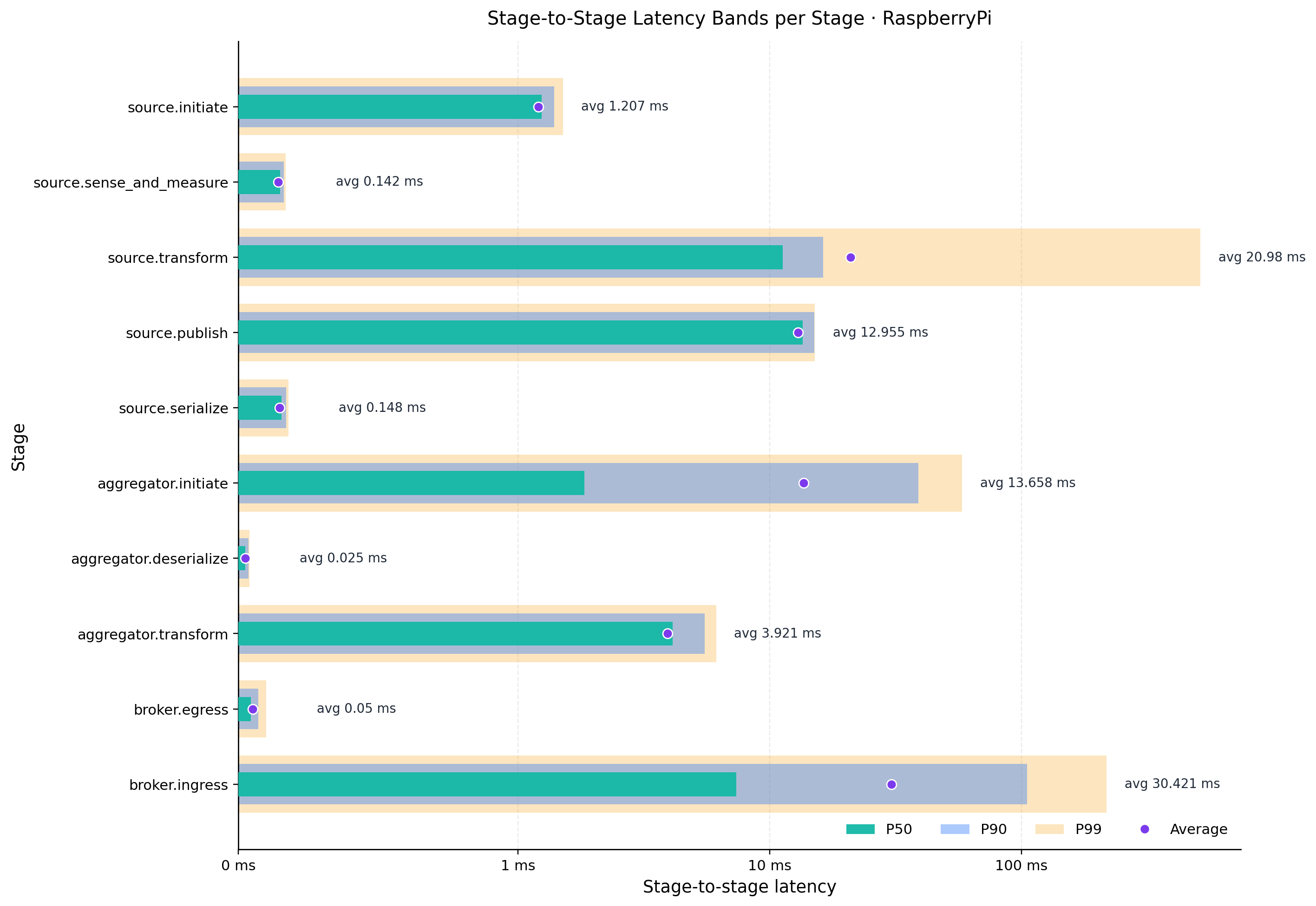}
  \caption{Stage-to-stage latency breakdown for RaspberryPi telemetry. Mean bars are overlaid with p50/p90/p99 markers (log-scale axis), enabling identification of consistently slow stages and stages with rare but substantial delays.}
  \label{fig:stage-breakdown}
\end{figure}

To localize where latency accumulates, the dataset supports stage-to-stage decomposition by taking differences between consecutive \emph{stage event timestamps} recorded in the telemetry envelope. In the current release, \texttt{stages\_long} stores one timestamp per stage key (event-style, without explicit start/end pairs); therefore, stage latencies are interpreted as inter-stage intervals along the fixed pipeline order. Figure~\ref{fig:stage-breakdown} shows the RaspberryPi breakdown, combining (i) a mean bar per stage with (ii) p50, p90, and p99 markers. The horizontal axis is displayed on a logarithmic scale to allow simultaneous inspection of sub-millisecond stages and tens-to-hundreds of millisecond stages.

\begin{table}[htbp]
\centering
\setlength{\tabcolsep}{6pt}
\begin{tabular}{lrrrr}
\toprule
Stage & avg (ms) & p50 (ms) & p90 (ms) & p99 (ms) \\
\midrule
broker.ingress            & 30.421019 &  7.364880 & 105.195288 & 218.066271 \\
source.transform          & 20.979565 & 11.258572 &  16.321526 & 514.414953 \\
aggregator.initiate       & 13.657719 &  1.832809 &  38.949239 &  58.020694 \\
source.publish            & 12.955387 & 13.515166 &  15.025323 &  15.131969 \\
aggregator.transform      &  3.921368 &  4.119082 &   5.510954 &   6.122134 \\
source.initiate           &  1.206680 &  1.241552 &   1.391939 &   1.506338 \\
source.serialize          &  0.147565 &  0.154183 &   0.170072 &   0.178571 \\
source.sense\_and\_measure &  0.142329 &  0.149628 &   0.162386 &   0.168350 \\
broker.egress             &  0.050283 &  0.043740 &   0.070240 &   0.099795 \\
aggregator.deserialize    &  0.024911 &  0.024203 &   0.035648 &   0.039462 \\
\bottomrule
\end{tabular}
\caption{Stage-to-stage latency summary for RaspberryPi messages (visualized in Figure~\ref{fig:stage-breakdown}).}
\label{tab:stage-breakdown}
\end{table}

\textbf{Observed characteristics.} The largest mean inter-event interval is the one ending at \texttt{broker.ingress} (30.42~ms), followed by \texttt{source.transform} (20.98~ms) and \texttt{aggregator.initiate} (13.66~ms). Several recorded intervals fall in the sub-millisecond regime (e.g., \texttt{aggregator.deserialize} at 0.025~ms mean). Heavy-tail behavior is visible for specific intervals; for example, \texttt{source.transform} has a p99 of 514.41~ms despite a mean near 21~ms, indicating rare but large delays relative to its typical behavior.

\subsubsection{Daily energy signature}
\label{subsubsec:daily-energy}

\begin{figure}[htbp]
  \centering
  \includegraphics[width=0.99\textwidth]{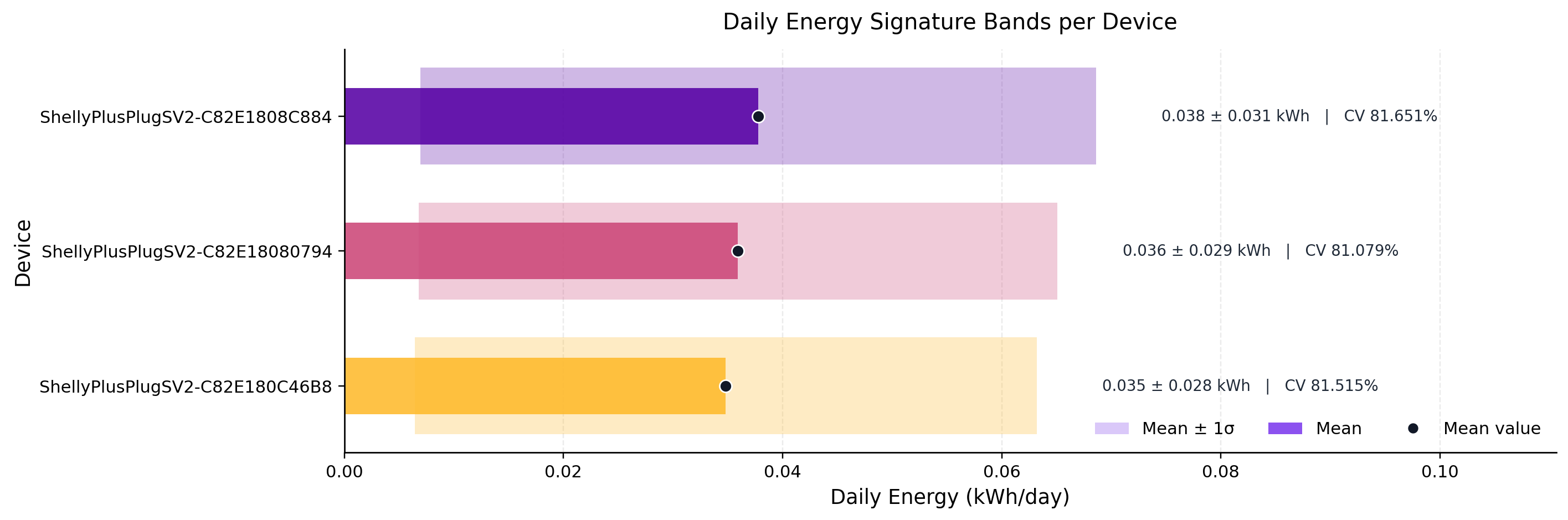}
  \caption{Daily energy signature per device: mean daily energy (kWh/day) with $\pm 1$ standard deviation whiskers and CV\% annotation. This view supports quick comparison of typical consumption and day-to-day variability.}
  \label{fig:daily-energy}
\end{figure}

Energy measurements captured by smart plugs enable daily summaries of per-node energy consumption. Figure~\ref{fig:daily-energy} reports each device's mean daily energy (kWh/day) with $\pm 1$ standard deviation whiskers, alongside the coefficient of variation (CV\%) as a normalized variability indicator.

\begin{table}[htbp]
\centering
\setlength{\tabcolsep}{6pt}
\begin{tabular}{lrrr}
\toprule
device\_id & mean\_kwh & std\_kwh & cv\_percent \\
\midrule
ShellyPlusPlugSV2-C82E18080794 & 0.035928 & 0.029130 & 81.078745 \\
ShellyPlusPlugSV2-C82E180C46B8 & 0.034814 & 0.028378 & 81.514856 \\
ShellyPlusPlugSV2-C82E1808C884 & 0.037763 & 0.030834 & 81.651410 \\
\bottomrule
\end{tabular}
\caption{Per-device daily energy summary (visualized in Figure~\ref{fig:daily-energy}).}
\label{tab:daily-energy}
\end{table}

\textbf{Observed characteristics.} The three \texttt{ShellyPlusPlugSV2} devices exhibit closely clustered mean daily energy values (approximately 0.0348--0.0378~kWh/day) and similar standard deviations (approximately 0.028--0.031~kWh/day), yielding high relative variability with CV near 81\% across devices (Table~\ref{tab:daily-energy}). These statistics summarize the observed calendar-day aggregates, including partial days, and are included as an example of the energy view supported by the testbed. They should not be interpreted as normalized full-day consumption estimates; differences in observation coverage can contribute to the reported variability.

\subsubsection{Energy versus data volume}
\label{subsubsec:energy-vs-volume}

\begin{figure}[htbp]
  \centering
  \includegraphics[width=0.99\textwidth]{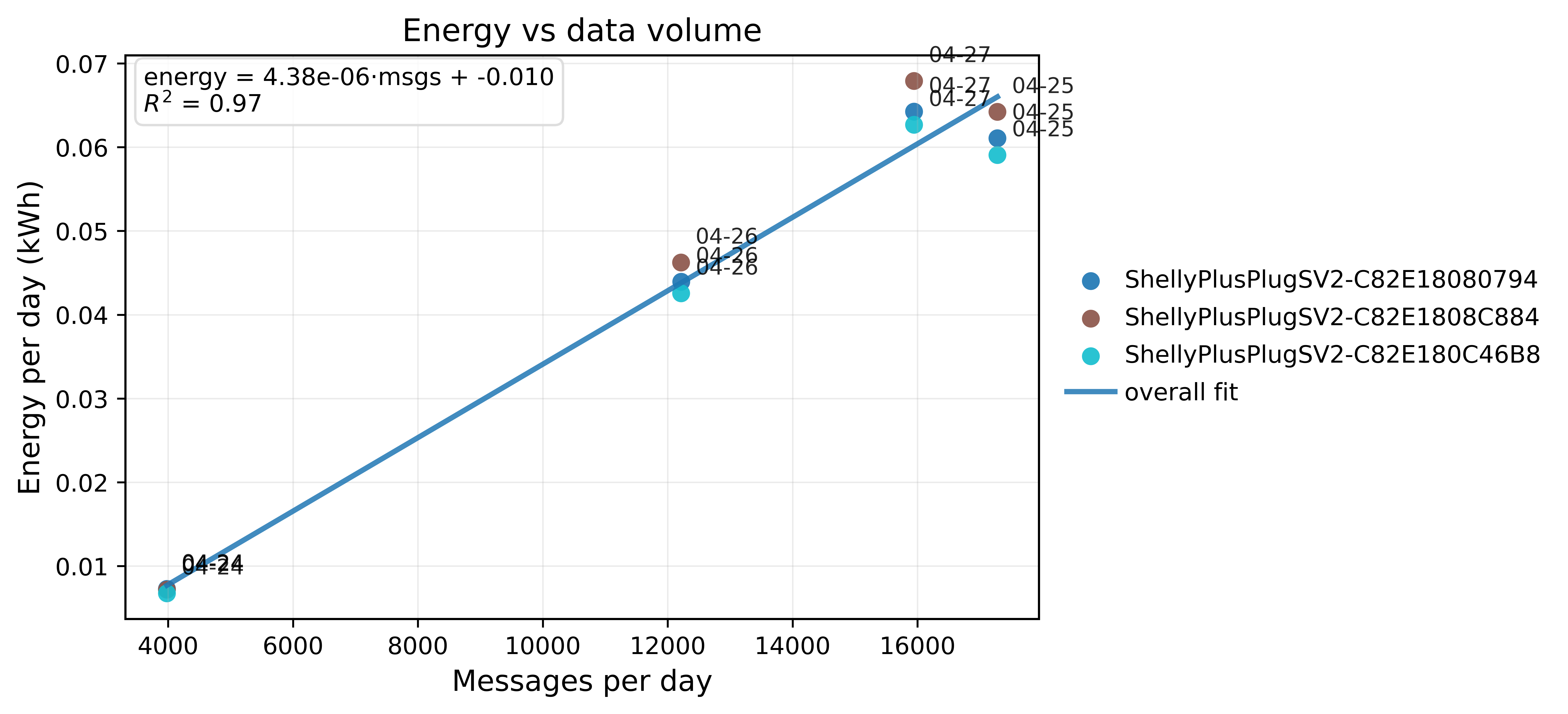}
  \caption{Energy versus data volume: daily energy (kWh/day) plotted against daily message count, with points grouped by device and a global linear fit. This visualization is intended for exploratory inspection of co-variation patterns in the demonstrated interval.}
  \label{fig:energy-vs-volume}
\end{figure}

A final exploratory check relates daily energy consumption to the volume of telemetry produced. Figure~\ref{fig:energy-vs-volume} plots energy (kWh/day) against daily message counts, grouping points by device and including a global linear fit to summarize the association in the observed interval.

\begin{table}[htbp]
\centering
\setlength{\tabcolsep}{6pt}
\begin{tabular}{llrr}
\toprule
device\_id & day & energy\_kwh\_day & msg\_count\_day \\
\midrule
ShellyPlusPlugSV2-C82E18080794 & 2025-04-24 & 0.007273 &  3{,}981 \\
ShellyPlusPlugSV2-C82E18080794 & 2025-04-25 & 0.061064 & 17{,}280 \\
ShellyPlusPlugSV2-C82E18080794 & 2025-04-26 & 0.043932 & 12{,}217 \\
ShellyPlusPlugSV2-C82E18080794 & 2025-04-27 & 0.064239 & 15{,}945 \\
ShellyPlusPlugSV2-C82E18080794 & 2025-04-28 & 0.003131 & \textit{NA} \\
\midrule
ShellyPlusPlugSV2-C82E1808C884 & 2025-04-24 & 0.007176 &  3{,}981 \\
ShellyPlusPlugSV2-C82E1808C884 & 2025-04-25 & 0.064214 & 17{,}280 \\
ShellyPlusPlugSV2-C82E1808C884 & 2025-04-26 & 0.046230 & 12{,}214 \\
ShellyPlusPlugSV2-C82E1808C884 & 2025-04-27 & 0.067899 & 15{,}945 \\
ShellyPlusPlugSV2-C82E1808C884 & 2025-04-28 & 0.003297 & \textit{NA} \\
\midrule
ShellyPlusPlugSV2-C82E180C46B8 & 2025-04-24 & 0.006745 &  3{,}981 \\
ShellyPlusPlugSV2-C82E180C46B8 & 2025-04-25 & 0.059059 & 17{,}280 \\
ShellyPlusPlugSV2-C82E180C46B8 & 2025-04-26 & 0.042555 & 12{,}215 \\
ShellyPlusPlugSV2-C82E180C46B8 & 2025-04-27 & 0.062670 & 15{,}945 \\
ShellyPlusPlugSV2-C82E180C46B8 & 2025-04-28 & 0.003038 & \textit{NA} \\
\bottomrule
\end{tabular}
\caption{Device-day pairs used in the energy-versus-volume visualization (Figure~\ref{fig:energy-vs-volume}).}
\label{tab:energy-vs-volume}
\end{table}

\textbf{Observed characteristics.} Over the displayed days (2025-04-24 to 2025-04-28), daily energy values range approximately from 0.003 to 0.068~kWh/day, while message counts include values such as 3{,}981, 12{,}214--12{,}217, 15{,}945, and 17{,}280 depending on device and day (Table~\ref{tab:energy-vs-volume}). The day 2025-04-28 contains missing message counts while still recording small energy values, indicating an incomplete pairing between the daily aggregates for that date in the demonstrated export. The linear fit is descriptive only: longer observation windows can increase both accumulated energy and message count, and background node consumption is not separated from telemetry-related work. Consequently, this association does not estimate marginal energy per message or demonstrate that message production causes the observed energy variation.

\section{Discussion}
\label{section:discussion}

This work presents an edge-focused IoT testbed and dataset whose primary practical contribution is the systematic co-capture of (i) heterogeneous telemetry (environmental sensing, OS/node monitoring, and plug-level energy metering) and (ii) stage-level timing metadata spanning Source, Broker, and Aggregator execution. Rather than proposing a new protocol, attribution method, or provenance mechanism, the testbed operationalizes a measurement discipline in which timeliness, energy, and provenance-related identifiers are treated as first-class data. The resulting dataset supports reproducible inspection of both payload values and the data-production pipeline that generated them.

\subsection{Implications of the observed dataset characteristics}

\paragraph{End-to-end latency differs by device class, with substantial tail behavior.}
The reported end-to-end latency distributions show clear separation between device types: RaspberryPi messages exhibit a mean latency of 83.51\,ms and median of 60.66\,ms, while ShellyPlusPlugSV2 messages exhibit a mean of 227.38\,ms and median of 155.47\,ms (Table~\ref{tab:e2e-latency}). At the same time, the upper tail is non-negligible for both classes, with p99 exceeding 500\,ms. This combination is practically important: median differences are relevant for typical responsiveness, whereas p95--p99 behavior is often decisive for systems with timeliness constraints, particularly when telemetry drives alerting or actuation decisions.

A key benefit of the proposed instrumentation is that such tail behavior is not only observable but can be analyzed in relation to pipeline stages. This provides candidate intervals for follow-up diagnosis, but does not identify a root cause or isolate persistence cost without the corresponding boundary measurements. Device-class differences also reflect different software, sensing workloads, and timestamp behavior; they should not be read as a controlled comparison of hardware or MQTT performance.

\paragraph{Stage-level decomposition localizes where latency accumulates and where variance concentrates.}
The stage-to-stage breakdown for RaspberryPi telemetry indicates that the largest mean intervals are associated with \texttt{broker.ingress} (30.42\,ms), \texttt{source.transform} (20.98\,ms), and \texttt{aggregator.initiate} (13.66\,ms), whereas several recorded intervals fall in the sub-millisecond regime (e.g., \texttt{aggregator.deserialize} at 0.025\,ms mean; Table~\ref{tab:stage-breakdown}). These values describe intervals between recorded events; they do not independently measure handler execution time or establish the overhead of serialization, deserialization, or instrumentation.

The distributional markers also highlight that variability is stage-dependent. For example, \texttt{source.transform} exhibits a p99 of 514.41\,ms despite a mean near 21\,ms, indicating rare but substantial increases in the interval ending at that event, without isolating transformation execution time. While the present paper does not attribute these events to specific causes, the recorded timestamps enable targeted follow-up studies (e.g., correlating stage delays with CPU load, scheduler effects, transient I/O, or network contention) without modifying the message format or the collection pipeline.

\paragraph{Measurement-context skew is substantial and should be treated as an analytic factor.}
The flattened representation reveals a strongly right-skewed context distribution: \texttt{OS} contributes 63.4\% of samples, while energy-related contexts (\texttt{energy}, \texttt{os}, \texttt{power}) also contribute materially, and environmental sensor contexts appear at lower frequencies (Table~\ref{tab:context-counts}). This is expected in heterogeneous IoT deployments where system telemetry has higher metric cardinality and stable cadence relative to many physical sensors.

Practically, this skew affects downstream analysis choices. Without explicit attention, learning or summarization procedures may primarily model OS and energy contexts due to volume, even when the application interest is environmental sensing. The long-format representation used here makes such imbalance explicit and supports mitigation through filtering, stratification, or reweighting. Additionally, the intentional distinction between \texttt{OS} (Raspberry Pi) and \texttt{os} (Shelly) encodes a platform boundary in the data, reducing the risk of accidental aggregation across non-equivalent telemetry sources.

\paragraph{Energy measurements provide a complementary operational view but require careful interpretation.}
Daily energy summaries for the three Shelly devices are closely clustered in mean consumption (approximately 0.0348--0.0378\,kWh/day), while exhibiting high relative variability (CV near 81\%; Table~\ref{tab:daily-energy}). These values are descriptive of the demonstrated capture window and illustrate variation among the observed daily aggregates, whose coverage is unequal. Several factors may contribute, including partial-day windows at the beginning or end of capture, changes in workload, and measurement/reporting granularity of smart plugs.

The energy-versus-volume exploration further shows that message counts align with the configured 5-second schedule (e.g., 17,280 messages/day), indicating that data volume is largely cadence-driven in this setup. However, the incomplete pairing observed for 2025-04-28 (energy recorded while message count is missing) emphasizes that daily aggregates should be interpreted in the context of collection completeness and window boundaries, and that the dataset publication pipeline should make such completeness signals easy to audit.

\subsection{Value of the dual representation and reproducible transformation}
A practical contribution of the testbed is the co-release of raw, schema-governed events (Avro) and curated, query-oriented tabular outputs (partitioned Parquet). The raw representation preserves message semantics and provides payloads for application-level replay, while the processed representation enables efficient exploratory analysis via predicate pushdown and time-based partitioning. The transformation is described as deterministic and streaming, and it is bidirectional in the sense that tabular rows can be reconstituted into TelemetryMessage objects. This reduces friction for practitioners who need to alternate between systems-level inspection (messages, topics, stages) and data-oriented analysis (windowing, joins, aggregates).

For reuse beyond this deployment, practitioners can use the released events as a fixed input corpus for testing decoders, schema handling, raw-to-tabular transformations, and analysis queries. The correlation identifiers allow a latency outlier selected from the stage table to be traced back to its readings and available message metadata. Replay into another MQTT pipeline can provide an input workload, but performance evaluation then requires a defined replay rate, fresh timestamps, and measurements from that target system; the original timings and energy values remain observations of this deployment. These are intended reuse paths, not demonstrated debugging case studies or evidence that the capture represents other deployment conditions.

\subsection{Limitations and threats to validity}

The current demonstration is intentionally scoped and should be interpreted accordingly.

Interpreting the reported quantities requires linking each measurement to the conditions under which it is produced. End-to-end and stage-to-stage latency estimates depend on clock alignment, timestamp granularity, and the precise semantics of stage timestamp keys; energy summaries depend on metering scope and window completeness; and provenance-oriented analyses depend on stable identifier conventions and consistent role instrumentation. The following threats to validity therefore map directly to the three measured dimensions emphasized in this work (timeliness, energy, and pipeline observability/provenance) and delimit how the demonstrated results should be generalized.

\begin{itemize}
  \item \textbf{Scale and topology.} Results reflect a small number of devices (three Raspberry Pi nodes and three smart plugs) operating on a single local gateway and broker. Latency and variability may differ under larger deployments, different Wi-Fi conditions, interference, multi-hop links, or WAN/cloud paths. Mobility, intermittent connectivity, battery operation, and failure recovery are not evaluated.
  \item \textbf{Clock synchronization.} NTP with a local reference provides clock alignment, but independent measurements of residual offsets and platform-specific timestamp granularity are not reported. Cross-node intervals therefore have unquantified timing uncertainty, limiting fine-grained comparisons and interpretation of both central and tail estimates.
  \item \textbf{Stage timestamp semantics.} In the processed view, stages are represented as event timestamps rather than explicit start/end pairs. Consequently, stage intervals are inferred as inter-event differences along a fixed pipeline order. These intervals localize elapsed time between recorded boundaries but do not separate queueing, scheduling, transport, and handler execution. They also do not establish durable persistence latency.
  \item \textbf{Energy metering scope.} Smart-plug measurements capture upstream consumption inclusive of adapter losses and background node activity. They do not isolate component-, process-, or message-level energy contributions, and the gateway is outside this metered scope. Independent calibration of meter accuracy and reporting resolution is not reported in the demonstration; co-capture therefore supports temporal association rather than causal energy attribution.
  \item \textbf{Validation and overhead.} No controlled workload sweep, uninstrumented baseline, or quantitative comparison with another testbed is provided. Timestamping and broker annotation costs, CPU/network overhead, and storage/export costs are not quantified. The present summaries therefore do not validate scalability, low instrumentation overhead, or storage efficiency.
  \item \textbf{Provenance scope.} Device and correlation identifiers, context labels, and stage metadata provide operational traceability within the recorded pipeline. They do not constitute authenticated provenance, tamper evidence, or a complete causal audit trail. No industrial debugging or provenance-auditing case study is evaluated.
  \item \textbf{Partial windows and completeness.} The presence of boundary windows with lower sample counts and at least one day with missing message counts indicates that completeness metadata should be considered when interpreting daily aggregates and correlations.
\end{itemize}

\subsection{Directions for extension}
The testbed is designed to support incremental extension rather than redesign. Several concrete enhancements would strengthen its utility for benchmarking and comparative studies: independent clock-error and meter calibration, instrumented-versus-uninstrumented overhead measurements, and a diagnosis case study with a known fault would help assess measurement validity and practical diagnostic value. Further extensions include (i) recording explicit start/end timestamps per stage to improve attribution of delays; (ii) varying message rate, payload size, MQTT QoS, subscriber count, broker load, CPU contention, and network loss under controlled conditions to characterize how distributions shift; (iii) expanding protocol and power diversity (e.g., BLE/Zigbee/Thread, battery-powered nodes) to evaluate generality; and (iv) propagating the same timeliness instrumentation into fog/cloud ingestion paths for end-to-end accounting beyond the edge.

\section{Conclusions}
\label{section:conclusions}

This paper presents an edge-focused IoT testbed and dataset that treat the telemetry pipeline itself as a measurable object. The contribution is organized around three practical axes: \emph{pipeline observability} via role-spanning stage timestamps across Source, Broker, and Aggregator; \emph{energy alignment} via plug-level metering co-captured with telemetry production; and a \emph{dual representation} that preserves both replayable raw events (Avro) and analysis-ready tabular views (partitioned Parquet) produced by a deterministic transformation workflow.

The demonstrated dataset shows that these choices enable immediate, reproducible inspection of system behavior beyond value-centric sensing traces: clear device-class separation in end-to-end latency with non-trivial tail behavior, stage-dependent variability that localizes where delay accumulates, pronounced measurement-context skew that must be treated as an analytic factor, and an operational energy view that supports day-level summaries and co-variation checks with data volume. These observations characterize the released capture rather than validate a general latency or energy model. By co-releasing payload telemetry together with timing and energy metadata in interoperable formats, the artifact provides a reusable input corpus and analysis structure for subsequent benchmarking and diagnostic studies. Controlled workload experiments, calibrated timing, overhead measurements, and concrete diagnosis cases remain necessary to substantiate broader claims.

Future extensions that directly strengthen this measurement discipline include recording explicit start/end timestamps per stage, expanding protocol and power diversity (e.g., battery-powered nodes and additional wireless stacks), and introducing controlled stress conditions to characterize distributional shifts under contention and loss.


\section*{Code and Data Availability}

The source code for the testbed implementation is publicly available at: \\ 
\url{https://github.com/gkoulis/tiny-testbed}

The dataset described in this paper is publicly available at: \\ 
\url{https://github.com/gkoulis/tiny-testbed-dataset}

The Eclipse Mosquitto plugin is publicly available at: \\ \url{https://github.com/gkoulis/mqtt-message-timestamps}

\bibliographystyle{unsrt}  
\bibliography{references}  

\end{document}